\documentclass[aps,prb,twocolumn,superscriptaddress,floatfix,10pt]{revtex4-1}
\usepackage[utf8]{inputenc}

\usepackage{natbib}
\usepackage{graphicx}
\usepackage{latexsym}
\usepackage{amssymb}
\usepackage{amsmath}
\usepackage{amsfonts}
\usepackage{amsthm}
\usepackage{bm}
\usepackage{enumitem}
\usepackage{hyperref}
\usepackage{lipsum}
\usepackage{braket}
\usepackage{tikz}
\usepackage{pgfplots}
\usepackage{comment}
\usepackage{subcaption}
\usepackage{ifthen}
\usepackage{caption}
\usepackage{ragged2e}
\usepackage{multirow}
\usetikzlibrary{arrows,decorations.pathreplacing,decorations.markings,arrows.meta,patterns,3d}
\usepackage[normalem]{ulem} 

\theoremstyle{definition}
\newtheorem{proposition}{Proposition}
\newtheorem{lemma}[proposition]{Lemma}
\newtheorem{thm}[proposition]{Theorem}
\newtheorem{protocol}[proposition]{Protocol}
\newtheorem{definition}[proposition]{Definition}

\hypersetup{
  colorlinks = true,
  urlcolor = blue,
  pdfauthor = {Xiuqi Ma, Wilbur Shirley, Meng Cheng, Michael Levin, John McGreevy, Xie Chen},
  pdftitle = {Ma et al. - 2020 - Fractonic order in infinite-component Chern-Simons gauge theories}
}

\begin{document}

\title{Ground state degeneracy of the Ising cage-net model}
\date{\today}

\author{Xiuqi Ma}
\affiliation{Department of Physics and Institute for Quantum Information and Matter, \mbox{California Institute of Technology, Pasadena, California 91125, USA}}

\author{Ananth Malladi}
\affiliation{\mbox{Department of Physics, University of California, Santa Barbara, CA 93106, USA}}

\author{Zongyuan Wang}
\affiliation{Department of Physics and Institute for Quantum Information and Matter, \mbox{California Institute of Technology, Pasadena, California 91125, USA}}

\author{Zhenghan Wang}
\affiliation{\mbox{Microsoft Station Q, Santa Barbara, California 93106-6105, USA}}
\affiliation{\mbox{Department of Mathematics, University of California, Santa Barbara, CA 93106, USA}}

\author{Xie Chen}
\affiliation{Department of Physics and Institute for Quantum Information and Matter, \mbox{California Institute of Technology, Pasadena, California 91125, USA}}

\begin{abstract} 
The Ising cage-net model, first proposed in \textit{Phys. Rev. X} 9, 021010 (2019), is a representative type I fracton model with nontrivial non-abelian features. In this paper, we calculate the ground state degeneracy of this model and find that, even though it follows a similar coupled layer structure as the X-cube model, the Ising cage-net model cannot be ``foliated" in the same sense as X-cube as defined in \textit{Phys. Rev. X} 8, 031051 (2018). A more generalized notion of ``foliation'' is hence needed to understand the renormalization group transformation of the Ising cage-net model. The calculation is done using an operator algebra approach that we develop in this paper, and we demonstrate its validity through a series of examples. 
\end{abstract}

\maketitle

\section{Introduction}
\label{sec:intro}

The Ising cage-net model (``Ising cage-net'' for short), which was first proposed in Ref.~\onlinecite{Abhinav2019}, is an interesting fracton model. Constructed by coupling intersecting stacks of $(2+1)$D doubled Ising string-net topological states via a mechanism called ``p-loop condensation'', the model was shown to host non-abelian fractional excitations that are free to move only along a line. This feature sharply distinguishes Ising cage-net from previously known fracton models like the X-cube model proposed in Ref.~\onlinecite{Vijay2016}, even though they are both of type I (meaning that not all fractional excitations are immobile and that the logical operators are non-fractal, unlike the Cubic Code proposed by Haah\cite{Haah2011}).

The X-cube model, and many of its abelian type I cousins, were found to have a nice property that we call ``foliation''\cite{Shirley2018}. That is, the system size of the model can be increased / decreased (between $L_x \times L_y \times L_z$ and $L_x \times L_y \times(L_z+1)$ for instance) by adding / removing decoupled $(2+1)$D topological layers and smooth deformation of the Hamiltonian near the layers. Here smooth deformation means slowly varying the Hamiltonian without closing the gap or, equivalently, applying a finite depth local unitary circuit. For the X-cube model, the topological layer involved is the $(2+1)$D toric code state. Many of the exotic properties of the X-cube model follow immediately from the foliation structure. For example, the ground state degeneracy grows by a factor of 4 when the system size grows by 1 in one direction. The planon excitations -- fractional excitations that are constrained to move only in a 2D plane -- are also inherited directly from the toric code layers. 

An interesting open question is whether Ising cage-net fits into this foliation scheme. Ising cage-net and the X-cube model are similar in many ways. In particular, both can be constructed using the coupled layer construction: X-cube by coupling toric code layers and Ising cage-net by coupling doubled Ising layers. It is hence tempting to guess that Ising cage-net is also foliated in the sense that its system size can be increased / decreased by adding / removing decoupled doubled Ising layers and smooth deformation. 

In this paper, we show that this cannot be the case. In particular, we calculate the ground state degeneracy (GSD) of Ising cage-net and observe that the GSD does not grow by integer factors when the system size increases in $x$, $y$ or $z$ directions. This contradicts with the foliation process where adding a decoupled 2D layer changes the GSD by an integer factor while smooth deformation keeps the GSD invariant. 

As an exactly solvable model, the GSD of Ising cage-net can in principle be explicitly determined, but the non-abelian nature of the model complicates the problem. We deal with it, as shown below, by focusing on the operator algebra of the logical operators in the ground space of the model. By logical operators, we mean all possible operators acting on the ground space. We use the fact that if the ground space has dimension $n$, then the logical operators form the algebra $\text{Mat}_n$ of all $n\times n$ complex matrices, and a maximal, abelian, diagonalizable subset among all logical operators (a Cartan subalgebra; Definition~\ref{def:Cartan}) has dimension $n$. Hence the dimension of the ground space can be determined from the dimension of either the full algebra or the Cartan subalgebra of the logical operators.

\begin{figure}[t]
    \centering
    \includegraphics[scale = 0.5]{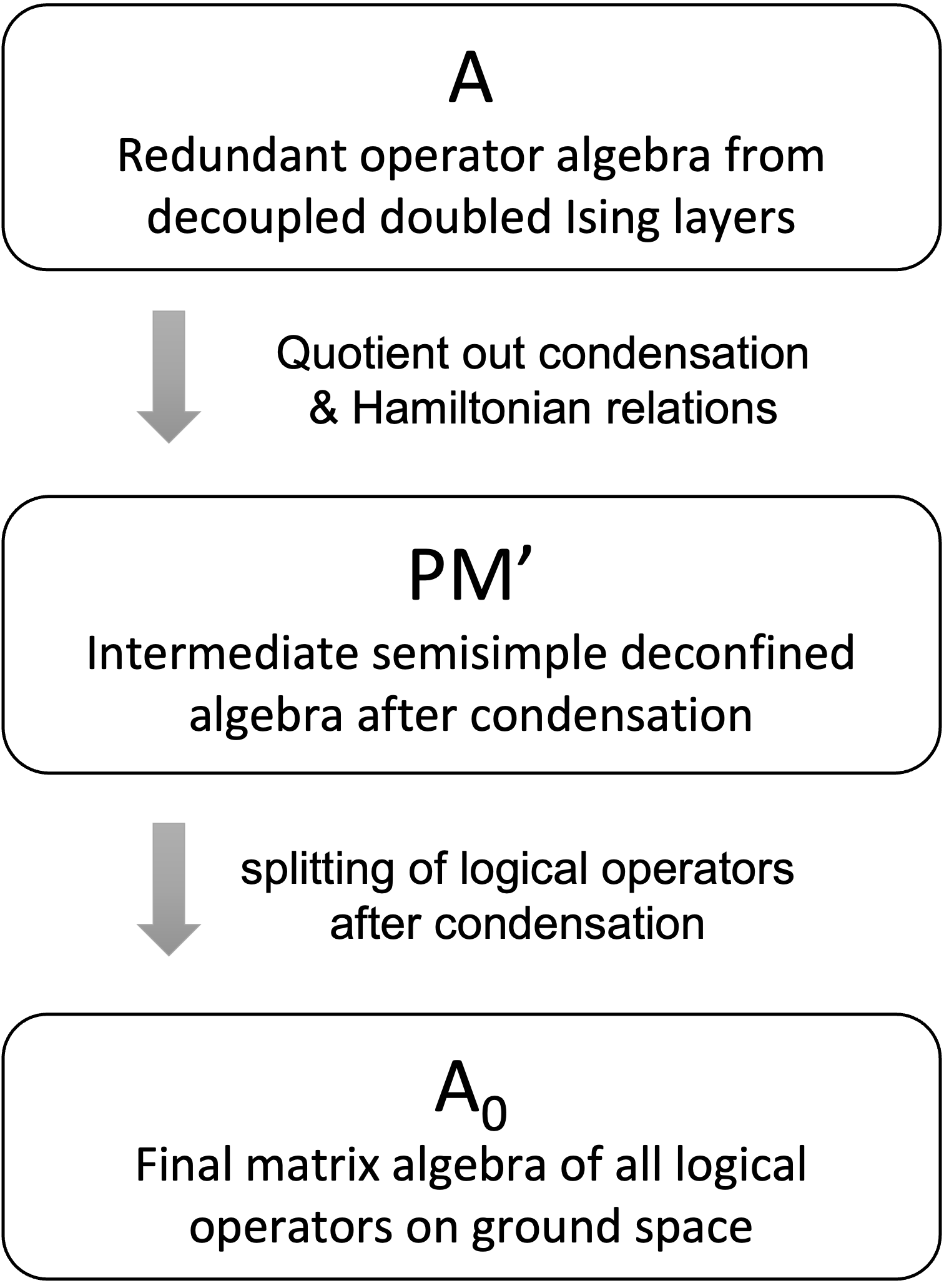}
    \captionsetup{justification=Justified}
    \caption{Procedure for determining the algebra of logical operators on the Ising cage-net ground space.}
    \label{fig:op_algebra}
\end{figure}

To determine the dimension of the operator algebras, we make use of the coupled layer condensation picture of the model. In particular, we start from a straight-forward but redundant operator algebra $A$ which is the tensor product of operators algebras  coming from the decoupled doubled Ising layers. The layers are coupled through $p$-loop condensation. Correspondingly, we take the commutant $M'$ of the condensate algebra $M$ within $A$, obtaining the deconfined algebra. Moreover, the deconfined algebra is further restricted by the constraints coming from the Hamiltonian (cube terms in Ising cage-net). When the set of both types of constraints is modded out, we get a semisimple algebra $PM'$, i.e.\ one with a block-diagonal form. The full logical operator algebra $A_0$ should be a matrix algebra, i.e.\ containing a single block. This structure will be recovered once we take into account the splitting of certain logical operators after condensation. This process is illustrated in Fig.~\ref{fig:op_algebra}. 

Following this procedure, we find the GSD of an $L_x\times L_y\times L_z$ Ising cage-net to be
\begin{equation}
   \text{GSD} = \frac{1}{8} \left(E_3 + E_2 + 5E_1 + 45\right),
   \label{eq:IsingCN_GSD_formula}
\end{equation}
where $E_3 = 9^{L_x+L_y+L_z}$, $E_2 = 9^{L_x+L_y}+9^{L_y+L_z}+9^{L_z+L_x}$, and $E_1 = 9^{L_x}+9^{L_y}+9^{L_z}$. Direct calculation shows that the GSD does not grow by integer multiples when the system size grows. For example, when $L_x=L_y=L_z=2$, $\text{GSD} =69048$; when $L_x=L_y=2,L_z=3$, $\text{GSD} = 614016$. Therefore, Ising cage-net cannot be foliated in the sense that its system size can be increased / decrease by adding / removing decoupled doubled Ising layers and smooth deformation.

In a separate paper\cite{https://doi.org/10.48550/arxiv.2301.00103}, we show that the system size of Ising cage-net can be changed
by ``condensing'' / ``uncondensing'' bosonic planon excitations near a 2D plane or, correspondingly,
through a linear depth circuit that scales with the size of the plane. This constitutes what we call the ``generalized foliated scheme'' for renormalization group transformation.

The paper is organized as follows: In Section~\ref{sec:H}, we review Ising cage-net. In Section~\ref{sec:chiral}, we introduce the operator algebra approach to calculating GSD by studying the simple example of the chiral Ising anyon model. The underlying mathematics of the operator algebra approach is the theory of semisimple algebras, and we discuss the structure of semisimple algebras in Section~\ref{sec:structure}. More mathematical details can be found in Appendix~\ref{app:math}. The construction of Ising cage-net involves p-loop condensation, and we study boson condensation in the operator algebra approach in Section~\ref{sec:SN} with the example of a condensation transition in the doubled Ising string-net model. We then use the operator algebra approach in Section~\ref{sec:1f-ICN} to study a more complicated $(2+1)$D topological order, the one-foliated Ising cage-net model. This model is closely related to Ising cage-net -- the main focus of this paper -- but is still a $(2+1)$D model so we can check the consistency of the operator algebra approach with anyon counting. Also in Section~\ref{sec:1f_Cartan}, we present another method of computing the GSD using a Cartan subalgebra. In Section~\ref{sec:ICN}, we put all of these tools together and compute the GSD of Ising cage-net in two ways. The correctness of our result \eqref{eq:IsingCN_GSD_formula} is further confirmed in Appendix~\ref{app:small_size} with lattice calculation for the smallest system size. Finally in Section~\ref{sec:summary}, we summarize our results and present several open questions.

\section{Review of the model}
\label{sec:H}

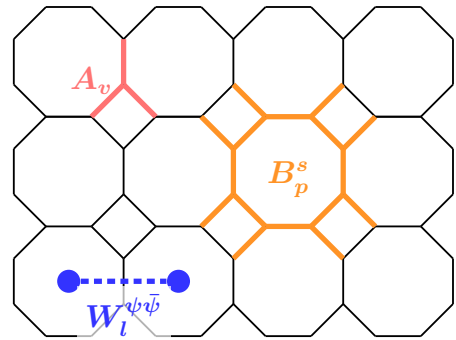
\begin{figure}[t]
    \centering
    \begin{tikzpicture}[baseline={([yshift=-.5ex]current bounding box.center)}, every node/.style={scale=1}]
        \pgfmathsetmacro{\lineW}{0.7}
        \pgfmathsetmacro{\lineWH}{2}
        \pgfmathsetmacro{\markingPos}{0.7}
        \pgfmathsetmacro{\radius}{0.79}
        \pgfmathsetmacro{\angle}{45}
        \pgfmathsetmacro{\startangle}{-22.5}
        \pgfmathsetmacro{\SquareL}{2*\radius*cos(22.5)}
        
        \begin{scope}[decoration={markings, mark=at position \markingPos with {\arrow{Latex}}}]
        \foreach \ix in {0,1,2,3}{
            \foreach \iy in {0,1,2} {
            
            \pgfmathsetmacro{\centerx}{\ix*\SquareL}
            \pgfmathsetmacro{\centery}{\iy*\SquareL}
            
                \foreach \i in {0,1,2,3,4,5,6,7} {
                    \pgfmathsetmacro{\ii}{\i+1}
                    \pgfmathsetmacro{\x}{\radius*sin(\startangle + \angle*\i) + \centerx}
                    \pgfmathsetmacro{\xx}{\radius*sin(\startangle + \angle*\ii) + \centerx}
                    \pgfmathsetmacro{\y}{\radius*cos(\startangle + \angle*\i) + \centery}
                    \pgfmathsetmacro{\yy}{\radius*cos(\startangle + \angle*\ii) + \centery}
                    
                    \ifthenelse{\ix = 3}{
                        \ifthenelse{\iy < 1} {
                                \ifthenelse{\i = 0}{}{
                                    \draw[line width = \lineW pt] (\xx , \yy) -- (\x , \y);
                                    }
                                }{
                                    \draw[line width = \lineW pt] (\xx , \yy) -- (\x , \y);
                                }
                    }{
                        \ifthenelse{\iy < 1}{
                            \ifthenelse{\i = 0 \OR \i = 2}{}{
                                    \draw[line width = \lineW pt] (\xx , \yy) -- (\x , \y);
                                    }
                            }{
                                \ifthenelse{\i = 2}{}{
                                    \draw[line width = \lineW pt] (\xx , \yy) -- (\x , \y);
                                    }
                            }
                    }
                }
            }
        }
        \end{scope}
        
        \begin{scope}
            \pgfmathsetmacro{\Lcenterx}{0}
            \pgfmathsetmacro{\Lcentery}{2*\SquareL}
            \pgfmathsetmacro{\Rcenterx}{\SquareL}
            \pgfmathsetmacro{\Rcentery}{2*\SquareL}
            \pgfmathsetmacro{\upptx}{\Lcenterx + \radius*sin(67.5)}
            \pgfmathsetmacro{\uppty}{\Lcentery + \radius*cos(67.5)}
            \pgfmathsetmacro{\centralptx}{\Lcenterx + \radius*sin(112.5)}
            \pgfmathsetmacro{\centralpty}{\Lcentery + \radius*cos(112.5)}
            \pgfmathsetmacro{\Leftptx}{\Lcenterx + \radius*sin(157.5)}
            \pgfmathsetmacro{\Leftpty}{\Lcentery + \radius*cos(157.5)}
            \pgfmathsetmacro{\Rightptx}{\Rcenterx + \radius*sin(202.5)}
            \pgfmathsetmacro{\Rightpty}{\Rcentery + \radius*cos(202.5)}
            
            \draw[line width = \lineWH pt, color = red!40!pink] (\upptx,\uppty) -- (\centralptx,\centralpty);
            \draw[line width = \lineWH pt, color = red!40!pink] (\Leftptx,\Leftpty) -- (\centralptx,\centralpty);
            \draw[line width = \lineWH pt, color = red!40!pink] (\Rightptx,\Rightpty) -- (\centralptx,\centralpty);
            
            \node [font = \large] at (\Leftptx, \Lcentery - 0.25) {$\bm{\textcolor{red!40!pink}{A_v}}$};
        \end{scope}
        
        \begin{scope}
        \pgfmathsetmacro{\centerx}{2*\SquareL}
        \pgfmathsetmacro{\centery}{\SquareL}
        
        \foreach \i in {0,1,2,3,4,5,6,7} {
            \pgfmathsetmacro{\ii}{\i+1}
            \pgfmathsetmacro{\x}{\centerx + \radius*sin(\startangle + \angle*\i)}
            \pgfmathsetmacro{\xx}{\centerx + \radius*sin(\startangle + \angle*\ii)}
            \pgfmathsetmacro{\y}{\centery + \radius*cos(\startangle + \angle*\i)}
            \pgfmathsetmacro{\yy}{\centery + \radius*cos(\startangle + \angle*\ii)}
            
            \draw[line width = \lineWH pt, color = orange!85!white] (\xx , \yy) -- (\x , \y);
            
            \pgfmathsetmacro{\sidelength}{2*\radius*sin(22.5)}
            \pgfmathsetmacro{\tmpxA}{\sidelength*sin(45+90*(\i-1)/2)}
            \pgfmathsetmacro{\tmpxB}{\sidelength*cos(45+90*(\i-1)/2)}
            
            \ifodd\i {
                \draw[line width = \lineWH pt, color = orange!85!white] (\xx + \tmpxA  , \yy + \tmpxB)--(\xx , \yy);
                \draw[line width = \lineWH pt, color = orange!85!white] (\x + \tmpxA  , \y + \tmpxB)--(\x , \y);
            }\fi
        }
        
        \node [font = \large] at (\centerx,\centery-0.1) {$\bm{\textcolor{orange!85!white}{B^s_p}}$};
    \end{scope}
    
    \begin{scope}
        \pgfmathsetmacro{\halfSquareL}{\SquareL/2}
        
        \draw (0,0) node[circle, fill = blue!80!white] (Left) {};
        \draw (\SquareL,0) node[circle, fill = blue!80!white] (Right) {};
        \draw[line width = 2 pt, dash pattern=on 3 pt off 2 pt, color = blue!80!white] (Left.east) -- (Right.west);
        
        \node [font = \large, color = blue!80!white, fill=white,opacity=0.7, text opacity=1] at (\halfSquareL,-0.45) {\bm{$W^{\psi\bar{\psi}}_l$}};
    \end{scope}
    \end{tikzpicture}
    \captionsetup{justification=Justified}
    \caption{A square-octagon lattice. A vertex term $A_v$ and a plaquette term $B^s_p$ are shown. The string operator $W^{\psi\bar{\psi}}_l$ creates a $\psi\bar{\psi}$ excitation on each of the two plaquettes bordering the edge $l$.}
    \label{fig:Square_octagon_lattice}
\end{figure}

The building block of Ising cage-net is the doubled Ising string-net model\cite{Stringnet} (``doubled Ising'' for short). As a string-net model, doubled Ising can be realized on any 2D trivalent lattice. For the purpose of constructing Ising cage-net later, we choose a square-octagon lattice (Fig.~\ref{fig:Square_octagon_lattice}). On each edge of the lattice, we put a local Hilbert space of dimension 3, with orthonormal basis vectors $\ket{0}$, $\ket{1}$ and $\ket{2}$. The labels $\{0,1,2\}$ are understood as values of ``strings'' located at the edges. We also need a set of symbols $(\delta_{ijk}, d_s, F^{ijm}_{kln})$, where all indices take values in $\{0,1,2\}$. For example, $\delta_{ijk}=1$ if $ijk=000$, 011, 022, 112 or their permutations, and $\delta_{ijk}=0$ otherwise.

The Hamiltonian consists of a vertex term $A_v$ for each vertex $v$ and a plaquette term $B_p$ for each plaquette $p$. The vertex term is \[
    A_v
    \Bigg \vert
    \begin{tikzpicture}[baseline={([yshift=-.5ex]current bounding box.center)}, every node/.style={scale=1}]
        \begin{scope}
            \pgfmathsetmacro{\lineW}{0.8}
            \pgfmathsetmacro{\radius}{0.7}
            \pgfmathsetmacro{\SquareL}{2*\radius*cos(22.5)}
            \pgfmathsetmacro{\Lcenterx}{0}
            \pgfmathsetmacro{\Lcentery}{2*\SquareL}
            \pgfmathsetmacro{\Rcenterx}{\SquareL}
            \pgfmathsetmacro{\Rcentery}{2*\SquareL}
            \pgfmathsetmacro{\upptx}{\Lcenterx + \radius*sin(67.5)}
            \pgfmathsetmacro{\uppty}{\Lcentery + \radius*cos(67.5)}
            \pgfmathsetmacro{\centralptx}{\Lcenterx + \radius*sin(112.5)}
            \pgfmathsetmacro{\centralpty}{\Lcentery + \radius*cos(112.5)}
            \pgfmathsetmacro{\Leftptx}{\Lcenterx + \radius*sin(157.5)}
            \pgfmathsetmacro{\Leftpty}{\Lcentery + \radius*cos(157.5)}
            \pgfmathsetmacro{\Rightptx}{\Rcenterx + \radius*sin(202.5)}
            \pgfmathsetmacro{\Rightpty}{\Rcentery + \radius*cos(202.5)}
            
            \draw[line width = \lineW pt] (\upptx,\uppty) -- (\centralptx,\centralpty);
            \draw[line width = \lineW pt] (\Leftptx,\Leftpty) -- (\centralptx,\centralpty);
            \draw[line width = \lineW pt] (\Rightptx,\Rightpty) -- (\centralptx,\centralpty);
            
            \node [] at (\Leftptx+0.2, \Lcentery+0.1) {$j$};
            \node [] at (\Leftptx, \Lcentery-0.38) {$i$};
            \node [] at (\Rightptx, \Lcentery-0.38) {$k$};
        \end{scope}
    \end{tikzpicture}
    \Bigg \rangle
    =
    \delta_{ijk}
    \Bigg \vert
    \begin{tikzpicture}[baseline={([yshift=-.5ex]current bounding box.center)}, every node/.style={scale=0.9}]
        \begin{scope}
            \pgfmathsetmacro{\lineW}{0.7}
            \pgfmathsetmacro{\radius}{0.7}
            \pgfmathsetmacro{\SquareL}{2*\radius*cos(22.5)}
            \pgfmathsetmacro{\Lcenterx}{0}
            \pgfmathsetmacro{\Lcentery}{2*\SquareL}
            \pgfmathsetmacro{\Rcenterx}{\SquareL}
            \pgfmathsetmacro{\Rcentery}{2*\SquareL}
            \pgfmathsetmacro{\upptx}{\Lcenterx + \radius*sin(67.5)}
            \pgfmathsetmacro{\uppty}{\Lcentery + \radius*cos(67.5)}
            \pgfmathsetmacro{\centralptx}{\Lcenterx + \radius*sin(112.5)}
            \pgfmathsetmacro{\centralpty}{\Lcentery + \radius*cos(112.5)}
            \pgfmathsetmacro{\Leftptx}{\Lcenterx + \radius*sin(157.5)}
            \pgfmathsetmacro{\Leftpty}{\Lcentery + \radius*cos(157.5)}
            \pgfmathsetmacro{\Rightptx}{\Rcenterx + \radius*sin(202.5)}
            \pgfmathsetmacro{\Rightpty}{\Rcentery + \radius*cos(202.5)}
            
            \draw[line width = \lineW pt] (\upptx,\uppty) -- (\centralptx,\centralpty);
            \draw[line width = \lineW pt] (\Leftptx,\Leftpty) -- (\centralptx,\centralpty);
            \draw[line width = \lineW pt] (\Rightptx,\Rightpty) -- (\centralptx,\centralpty);
            
            \node [] at (\Leftptx+0.2, \Lcentery+0.1) {$j$};
            \node [] at (\Leftptx, \Lcentery-0.38) {$i$};
            \node [] at (\Rightptx, \Lcentery-0.38) {$k$};
        \end{scope}
    \end{tikzpicture}
    \Bigg \rangle ,
\] which allows certain ways for the strings to ``fuse'' at a vertex at low energy. The plaquette term is \[B_p=\frac{\sum_s d_sB_p^s}{\sum_s d_s^2},\] where the operator $B_p^s$ involves the symbols $F^{ijm}_{kln}$ and essentially acts by fusing an $s$-loop into the plaquette $p$; the precise definition of $B_p^s$ is not important here. The full Hamiltonian is then \[H = - \sum_v A_v - \sum_p B_p.\] This is a commuting projector Hamiltonian when restricted to the low-energy subspace where $A_v=1$ for all $v$. It has anyons 1, $\sigma$, $\bar\sigma$, $\psi$, $\bar\psi$, $\sigma\bar\psi$, $\psi\bar\sigma$, $\sigma\bar\sigma$ and $\psi\bar\psi$, where $\bar\psi$ is the time-reversal of $\psi$ but otherwise unrelated to $\psi$, and similarly for $\bar\sigma$. In fact, doubled Ising can be viewed as the chiral Ising anyon model\cite{PhysRevLett.86.268} (more discussion in Section~\ref{sec:chiral}) which has anyons 1, $\sigma$ and $\psi$, stacked with its time-reversal which has anyons 1, $\bar\sigma$ and $\bar\psi$. This is where the name ``doubled'' Ising comes from. The fusion rules for $\sigma$ and $\psi$ are $\sigma\times\sigma=1+\psi$, $\sigma\times\psi=\sigma$, $\psi\times\psi=1$; similarly for $\bar\sigma$ and $\bar\psi$. The $R$-symbols and string operators of the anyons can be found in Ref.~\onlinecite{Stringnet}, and we mention some important ones here:
\begin{itemize}[nolistsep, leftmargin=*]
    \item The braiding of $\sigma$ with $\psi$ gives a phase $-1$, and $\psi$ braids trivially with $\psi$; same for $\bar\sigma$ and $\bar\psi$.
    
    \item The operator $W^{\psi\bar{\psi}}_{l} = (-1)^{n_1(l)}$ creates a $\psi\bar{\psi}$ excitation on each of the two plaquettes bordering the edge $l$, where $n_1(l) = 1$ if the state on the edge $l$ is $\ket{1}$, and $n_1(l) = 0$ otherwise (Fig.~\ref{fig:Square_octagon_lattice}). We can extend the blue dashed line to obtain a string operator of $\psi\bar\psi$.
\end{itemize}
As a $(2+1)$D topological order, the GSD of doubled Ising is equal to the number of anyons\cite{Wen:2019ylt}, i.e.\ $\text{GSD}=9$.

\begin{figure}[t]
    \centering
    \includegraphics[scale = 0.1]{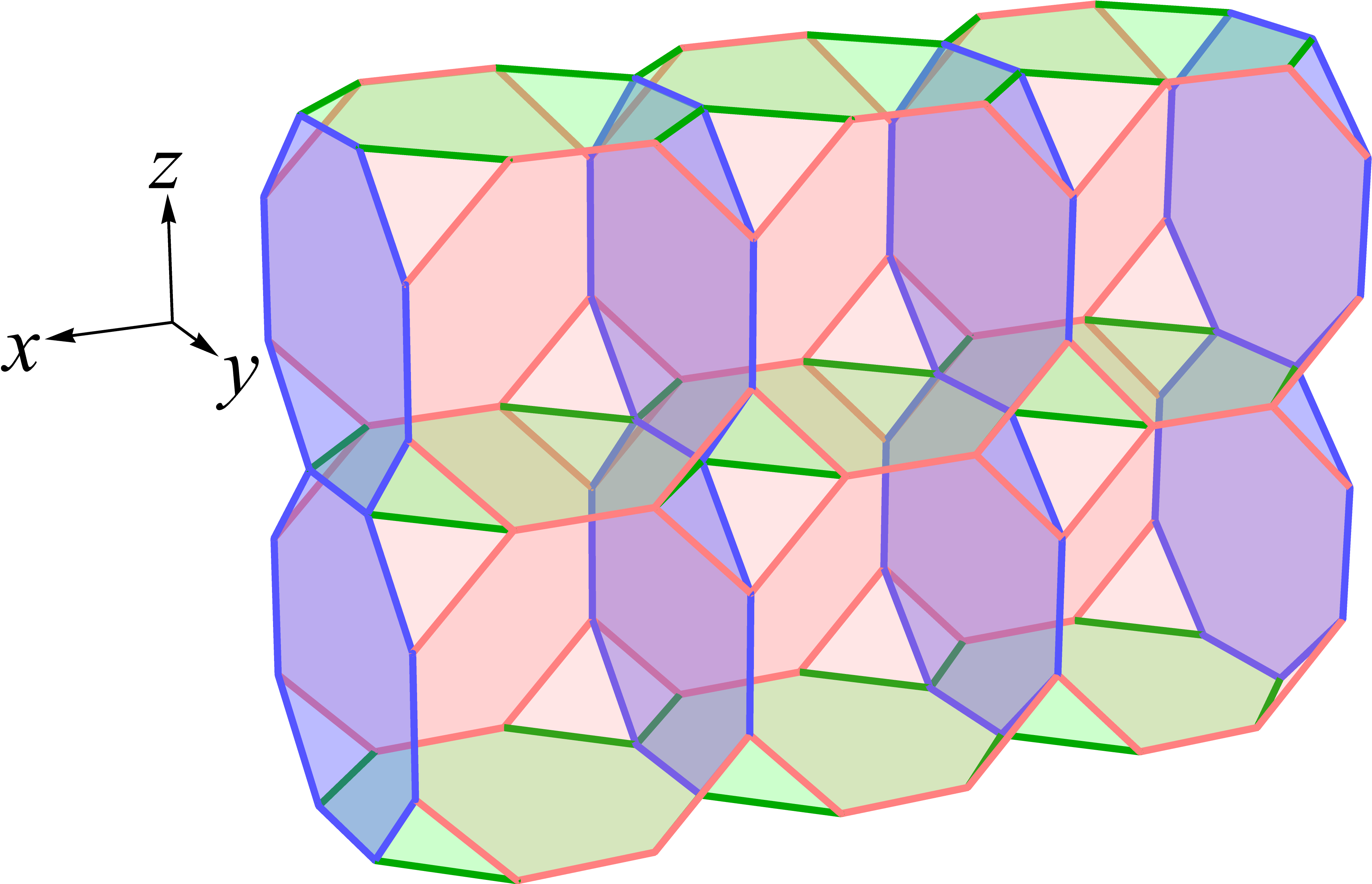}
    \captionsetup{justification=Justified}
    \caption{A truncated cubic lattice built from intersecting layers of the square-octagon lattice.}
    \label{fig:Truncated_cubic_lattice2}
\end{figure}

\begin{figure}[t]
    \centering
    \includegraphics[scale = 0.07]{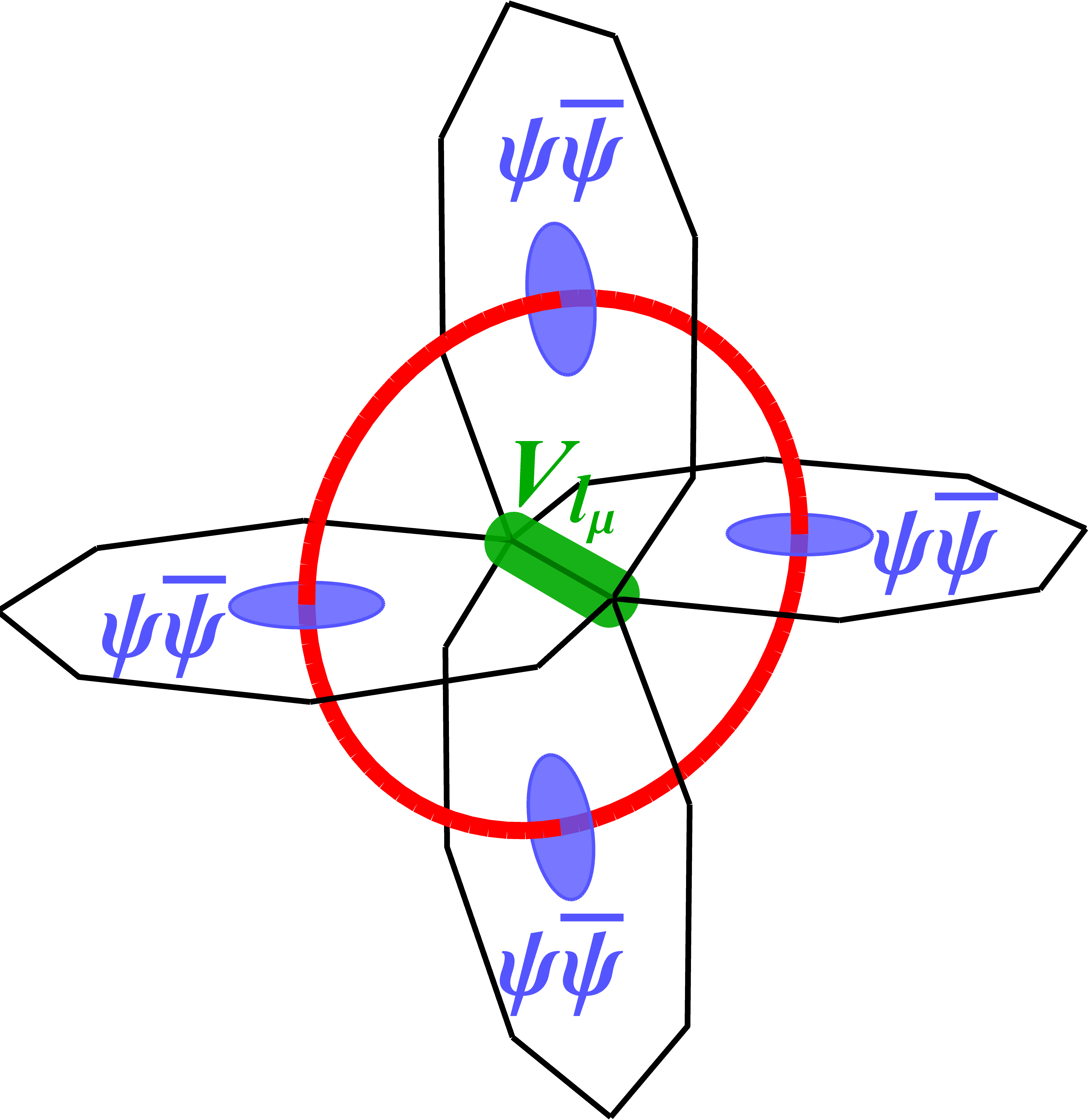}
    \captionsetup{justification=Justified}
    \caption{A $\psi\bar{\psi}$ p-loop shown in red. It is created by the operator $V_{l_\mu}$, the green cylinder. Connecting the $\psi\bar\psi$ particles with line segments orthogonal to their hosting plaquettes, we obtain the p-loop.}
    \label{fig:p-loop}
\end{figure}

\begin{figure}[t]
    \centering
    \includegraphics[scale = 0.1]{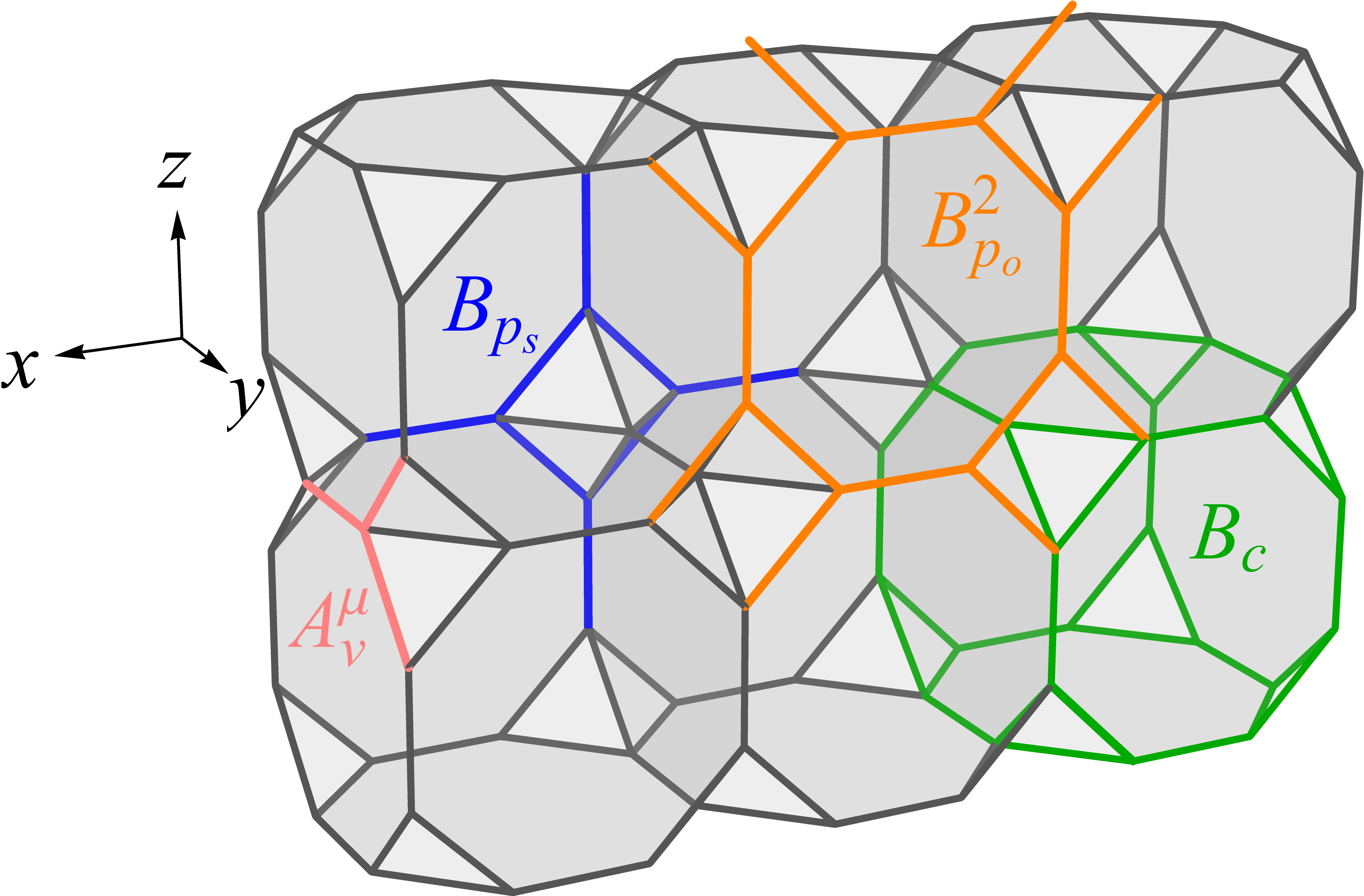}
    \captionsetup{justification=Justified}
    \caption{Hamiltonian terms of Ising cage-net. The full Hamiltonian is given by \eqref{eq:H}.}
    \label{fig:terms}
\end{figure}

To construct Ising cage-net\cite{Abhinav2019}, we first stack up layers of doubled Ising in the $x$, $y$ and $z$ directions. The resulting lattice is a truncated cubic lattice (Fig.~\ref{fig:Truncated_cubic_lattice2}). In this lattice, an edge $l_\mu$ parallel to the $\mu$ direction for $\mu=x$, $y$ or $z$ is called a principal edge. We will also distinguish the octagon and square plaquettes, denoting them by $p_\text{o}$ and $p_\text{s}$, respectively. On a principal edge $l_\mu$, the operator
\begin{equation}\label{eq:condense_op}
    V_{l_\mu} = W^{(\psi\bar{\psi})^\nu}_{l_\mu}W^{(\psi\bar{\psi})^\rho}_{l_\mu}=(-1)^{n_1^\nu(l_\mu)}(-1)^{n_1^\rho(l_\mu)}
\end{equation}
creates a $\psi\bar\psi$ particle-loop (``p-loop'' for short) around the edge (Fig.~\ref{fig:p-loop}), where $\mu$, $\nu$ and $\rho$ are distinct. To be precise, $a^\mu(i)$ denotes the anyon $a$ in the $i$th plane orthogonal to the $\mu$ direction, and we may omit the $i$ label when it is clear from context. For example, if $\mu=x$ then we can take $\nu=y$, $\rho=z$, and the $\psi\bar\psi$ particles in the p-loop originate from the $xz$ and $xy$ planes. We can condense these p-loops with the Hamiltonian \[H_0-J\sum_\mu\sum_{l_\mu}V_{l_\mu},\] where $H_0$ is the Hamiltonian for the decoupled layers of doubled Ising, and $J>0$ is a large coefficient enforcing the condensation. This reduces the low-energy Hilbert space on each edge to one of dimension 5, spanned by $\ket{00}$, $\ket{02}$, $\ket{20}$, $\ket{22}$ and $\ket{11}$. If we apply perturbation theory with $H_0$ as the perturbation, the plaquette terms $B_{p_\text{o}}^1$ must be assembled into cube terms \[B_c=\prod_{p_\text{o}\in c}\frac{\sqrt{2}}{4}B_{p_\text{o}}^1\] for each cube $c$. The resulting Hamiltonian of Ising cage-net is
\begin{equation}\label{eq:H}
    H=-\sum_{v,\mu} A_v^\mu-\sum_{p_\text{s}}B_{p_\text{s}}-\sum_{p_\text{o}}\frac{1}{4}(1+B_{p_\text{o}}^2)-\sum_c B_c,
\end{equation}
where $A_v^\mu$ is the vertex term at vertex $v$ orthogonal to the $\mu$ direction, and $B_{p_\text{o}}^2$ is the plaquette term of the 2-loop (not the square of an operator). The terms are shown in Fig.~\ref{fig:terms}. This is a commuting projector Hamiltonian when restricted to the low-energy subspace where all vertex terms are satisfied.

In order for an anyon to remain deconfined upon condensation, its string operator must commute with $V_{l_\mu}$. In other words, the anyon must braid trivially with the $\psi\bar\psi$ p-loop. For example, a $\sigma$ planon in an $xy$ plane has a braiding phase $-1$ with a $\psi\bar\psi$ p-loop created by some $V_{l_x}$ or $V_{l_y}$, and is therefore confined. On the other hand, a $\sigma$ planon in an $xy$ plane combines with a $\sigma$ planon in an $xz$ plane to form a lineon that moves in the $x$ direction, and this lineon is deconfined. We summarize the deconfined excitations in Table~\ref{table:Ising_CageNet_Excitations}.

\begin{table}[t]
\begin{center}
\begin{tabular}{c | c | c}
\toprule 
\multirow{1}{*}{Mobility}& \multirow{1}{*}{Type} & Excitations\\
\colrule
\multirow{3}{*}{Planon} & Abelian & $\psi^\mu(i)$, $\bar\psi^\mu(i)$\\
\cline{2-3}
 & \multirow{2}{*}{Non-abelian}  & $\sigma^\mu(i)\sigma^\mu(j)$, $\bar{\sigma}^\mu(i)\sigma^\mu(j)$,\\
& & $\sigma^\mu(i)\bar{\sigma}^\mu(j)$, $\bar{\sigma}^\mu(i)\bar{\sigma}^\mu(j)$\\
\hline
\multirow{3}{*}{Lineon} & Abelian &\\
\cline{2-3}
& \multirow{2}{*}{Non-abelian} & $\sigma^\mu(i)\sigma^\nu(j)$, $\bar{\sigma}^\mu(i)\sigma^\nu(j)$,\\
& & $\sigma^\mu(i)\bar{\sigma}^\nu(j)$, $\bar{\sigma}^\mu(i)\bar{\sigma}^\nu(j)$\\
\botrule
\end{tabular}
\captionsetup{justification=Justified}
\caption{Elementary excitations in Ising cage-net. We have $\mu\neq\nu$ in the lineon sector. The lineon $\sigma^x(i) \sigma^y(j)$ moves in the $z$ direction; similarly for the other non-abelian lineons.}
\label{table:Ising_CageNet_Excitations}
\end{center}
\end{table}

Although Ising cage-net is exactly solvable, it is not obvious how to calculate its GSD. In the following sections, we will introduce a new way of calculating the GSD that applies to Ising cage-net. We will start with some simple $(2+1)$D topological orders, and work our way up to Ising cage-net.

\section{GSD of chiral Ising}
\label{sec:chiral}

The chiral Ising anyon model\cite{PhysRevLett.86.268} (``chiral Ising'' for short) is a $(2+1)$D topological order whose properties such as GSD and anyon structure are well-known. As explained in Section~\ref{sec:H}, we can use chiral Ising to construct doubled Ising and hence Ising cage-net. In this section, we review chiral Ising and calculate its GSD with an alternative method, namely the operator algebra approach. While this approach may seem over-complicated for this relatively simple model, we aim to set up the general formalism and present several useful mathematical statements, as this approach will be used in Section~\ref{sec:ICN} for calculating the GSD of Ising cage-net.

There are three anyons in chiral Ising: 1, $\sigma$ and $\psi$. This model can be obtained e.g.\ by gauging the $\mathbb{Z}_2$ fermion parity symmetry in a $p+ip$ superconductor. In this context, 1 is the vacuum, $\sigma$ is the $\pi$ gauge flux, and $\psi$ is the gauge charge. The fusion rules are $\sigma\times\sigma=1+\psi$, $\sigma\times\psi=\sigma$, $\psi\times\psi=1$. The $F$- and $R$-symbols can be found in Ref.~\onlinecite{bonderson2012non}. The GSD of a $(2+1)$D topological order on a torus is equal to its number of anyons, so the Ising anyon model has $\text{GSD}=3$. This is equivalent to saying that the algebra of logical operators is $A_0=\text{Mat}_3$. Here, $\text{Mat}_n$ is the set of all $n\times n$ complex matrices. In the operator algebra approach which we will discuss now, we treat $A_0$ as the more fundamental object, design a framework to compute $A_0$ without knowledge of the ground space $\mathcal{H}_0$, and view $\mathcal{H}_0$ as a representation space of $A_0$.

The starting point of the operator algebra approach is a set of logical operators. We require these operators to linearly span the vector space of all logical operators, but we don't require them to be linearly independent. For a $(2+1)$D topological order on a torus, we take these to be operators of the form
\begin{equation}\label{eq:trivalent}
    v(a,b,c)=\vcenter{\hbox{
\begin{tikzpicture}
    \draw (-1, -1) rectangle (1, 1);
    \draw[->, -to] (-1, 0) arc (90:67.5:1);
    \draw (-0.0761, -0.6173) arc (22.5:67.5:1);
    \draw[->, -to] (0, -1) arc (0:22.5:1);
    \draw (1, 0) arc (270:247.5:1);
    \draw[->, -to] (0.2929, 0.2929) arc (225:247.5:1);
    \draw[->, -to] (0.2929, 0.2929) arc (225:202.5:1);
    \draw (0, 1) arc (180:202.5:1);
    \draw[->, -to] (-0.2929, -0.2929) -- (0, 0);
    \draw (0, 0) -- (0.2929, 0.2929);
    \node[above] at (-0.6173, -0.0761) {$a$};
    \node[right] at (-0.0761, -0.6173) {$b$};
    \node[below] at (0.6173, 0.0761) {$a$};
    \node[left] at (0.0761, 0.6173) {$b$};
    \node[below right=-2pt] at (0, 0) {$c$};
\end{tikzpicture}}}~,
\end{equation}
where $a, b, c$ are anyons consistent with the fusion rules (for simplicity we assume no fusion multiplicity). We call such an operator an \textit{elementary operator}. If $b=1$ then we must have $a=c$, and we will sometimes use the short-hand notation $a_x=v(a,1,a)$; similarly $b_y=v(1,b,b)$.

Of course, an elementary operator acts on the ground space $\mathcal{H}_0$ and has a matrix representation, but our discussion here does not rely on such a representation. Instead, we view the elementary operators as abstract objects. We can form a complex vector space $A$ over the elementary operators, with formal addition and formal scalar multiplication. The vector space $A$ has an additional operation called multiplication, defined for a pair of elementary operators by stacking one operator on top of the other and reducing the diagram to a sum of elementary operators using $F$- and $R$- symbols:
\begin{align*}
    v(a,b,c)v(a',b',c')&=\vcenter{\hbox{
\begin{tikzpicture}
    \draw (-1, -1) rectangle (1, 1);
    \draw (-1, -0.3333) arc (90:0:0.6667);
    \draw (1, 0.3333) arc (270:180:0.6667);
    \draw (-0.3333, 0.3333) arc (90:0:0.6667);
    \draw[->, -to] (-1, 0.3333) -- (-0.6667, 0.3333);
    \draw (-0.6667, 0.3333) -- (-0.3333, 0.3333);
    \draw[->, -to] (0.3333, -1) -- (0.3333, -0.6667);
    \draw (0.3333, -0.6667) -- (0.3333, -0.3333);
    \draw[line width=4pt, white] (-0.3333, 0.9) -- (-0.3333, 0.3);
    \draw[line width=4pt, white] (0.9, -0.3333) -- (0.3, -0.3333);
    \draw (-0.3333, 0.3333) arc (180:270:0.6667);
    \draw[->, -to] (-0.3333, 0.3333) -- (-0.3333, 0.6667);
    \draw (-0.3333, 0.6667) -- (-0.3333, 1);
    \draw[->, -to] (0.3333, -0.3333) -- (0.6667, -0.3333);
    \draw (0.6667, -0.3333) -- (1, -0.3333);
    \draw[->, -to] (-0.5286, -0.5286) -- (-0.3333,-0.3333);
    \draw (-0.3333,-0.3333) -- (-0.1381, -0.1381);
    \draw[->, -to] (0.1381, 0.1381) -- (0.3333, 0.3333);
    \draw (0.3333, 0.3333) -- (0.5286, 0.5286);
    \node[above] at (0.6667, -0.3333) {$a$};
    \node[left] at (-0.3333, 0.6667) {$b$};
    \node[below] at (-0.6667, 0.3333) {$a'$};
    \node[right] at (0.3333, -0.6667) {$b'$};
    \node[below right=-2pt] at (-0.3333, -0.3333) {$c$};
    \node[above left=-2.5pt] at (0.3333, 0.3333) {$c'$};
\end{tikzpicture}}}\\
&=\sum_{f,g}\sqrt{\frac{d_fd_g}{d_ad_{a'}d_bd_{b'}}}\vcenter{\hbox{
\begin{tikzpicture}
    \draw (-1, -1) rectangle (1, 1);
    \draw[->, -to] (-1, 0) -- (-0.8, 0);
    \draw (-0.8, 0) -- (-0.7, 0);
    \draw[->, -to] (0.7, 0) -- (0.9, 0);
    \draw (0.9, 0) -- (1, 0);
    \draw[->, -to] (0, -1) -- (0, -0.8);
    \draw (0, -0.8) -- (0, -0.7);
    \draw[->, -to] (0, 0.7) -- (0, 0.9);;
    \draw (0, 0.9) -- (0, 1);
    \draw (-0.7, 0) arc (180:90:0.25);
    \draw (0.45, -0.25) arc (-90:0:0.25);
    \draw (0, -0.7) arc (-90:0:0.25);
    \draw (-0.25, 0.45) arc (180:90:0.25);
    \draw[->, -to] (-0.45, 0.25) -- (-0.4, 0.25);
    \draw (-0.4, 0.25) -- (0, 0.25);
    \draw[->, -to] (0.25, -0.45) -- (0.25, -0.4);
    \draw (0.25, -0.4) -- (0.25, 0);
    \draw (0.25, 0) arc (0:90:0.25);
    \draw[line width=4pt, white] (-0.25, 0.2) -- (-0.25, 0.3);
    \draw[line width=4pt, white] (0.2, -0.25) -- (0.3, -0.25);
    \draw (-0.25, 0) arc (180:270:0.25);
    \draw[->, -to] (0, -0.25) -- (0.45, -0.25);
    \draw (0.45, -0.25) -- (0.5, -0.25);
    \draw[->, -to] (-0.25, 0) -- (-0.25, 0.45);
    \draw (-0.25, 0.45) -- (-0.25, 0.5);
    \draw[->, -to] (-0.7, 0) -- (-0.7, -0.35);
    \draw (-0.7, -0.35) -- (-0.7, -0.45);
    \draw[->, -to] (0, -0.7) -- (-0.35, -0.7);
    \draw (-0.35, -0.7) -- (-0.45, -0.7);
    \draw (-0.7, -0.45) arc (180:270:0.25);
    \draw[->, -to] (0.7, 0.45) -- (0.7, 0.25);
    \draw (0.7, 0.25) -- (0.7, 0);
    \draw[->, -to] (0.45, 0.7) -- (0.25, 0.7);
    \draw (0.25, 0.7) -- (0, 0.7);
    \draw (0.7, 0.45) arc (0:90:0.25);
    \draw[->, -to] (-0.6268, -0.6268) -- (-0.4018, -0.4018);
    \draw (-0.4018, -0.4018) -- (-0.1768, -0.1768);
    \draw[->, -to] (0.1768, 0.1768) -- (0.4018, 0.4018);
    \draw (0.4018, 0.4018) -- (0.6268, 0.6268);
    \node[below right=-3pt] at (-0.35, -0.35) {$c$};
    \node[below right=-4.5pt] at (0.35, 0.35) {$c'$};
    \node[right=-3pt] at (0.7, 0.5) {$a'$};
    \node[left=-3pt] at (-0.7, -0.45) {$a$};
    \node[below=-3pt] at (-0.5, -0.7) {$b$};
    \node[below=-2pt] at (0.15, -0.7) {$g$};
    \node[above=-2.5pt] at (0.5, 0.7) {$b'$};
    \node[above=-3pt] at (-0.85, 0.1) {$f$};
    \node[right=-2pt] at (0.2, -0.6) {$b'$};
    \node[above=-2pt] at (-0.55, 0.25) {$a'$};
    \node[above=-2pt] at (-0.3, 0.55) {$b$};
    \node[below=-2pt] at (0.65, -0.2) {$a$};
\end{tikzpicture}}}\\
&=\sum_{f,g,h}\lambda(f,g,h)v(f,g,h),
\end{align*}
with some coefficients $\lambda(f,g,h)$. Here $f$, $g$ and $h$ are some anyons, and $d_a$ is the quantum dimension of $a$. Going from the first line to the second line, we fused $a$ with $a'$ to get $f$, and $b$ with $b'$ to get $g$; going from the second line to the third line, we used $F$- and $R$-moves to transform the diagrams into elementary operators. In principle, we can compute $\lambda(f,g,h)$ for a general anyon theory, but in this paper we will only need some simple cases. For example, in chiral Ising we have \[\psi_x\psi_y=v(\psi,1,\psi)v(1,\psi,\psi)=-v(\psi,\psi,1),\] where the minus sign comes from $R_1^{\psi,\psi}=-1$. The multiplication has an identity $1=v(1,1,1)$. We say that $A$ is an \textit{algebra}, which is a complex vector space equipped with multiplication and a multiplicative identity (Definition~\ref{def:algebra} in Appendix~\ref{app:details} explains this concept more rigorously). If one views the elements of $A$ as operators on $\mathcal{H}_0$, then the addition, scalar multiplication and multiplication are the usual matrix operations. However, we stress again that we would like to view $A$ as a structure in its own right and not interpret it as a matrix algebra acting on a Hilbert space just yet.

For chiral Ising, we have 10 elementary operators
\[
\begin{aligned}
&v(1,1,1), &&v(\psi,1,\psi),\\
&v(1,\psi,\psi), &&v(\psi,\psi,1),\\
&v(\sigma,1,\sigma), &&v(1,\sigma,\sigma),\\
&v(\sigma,\psi,\sigma), &&v(\psi,\sigma,\sigma),\\
&v(\sigma,\sigma,1), &&v(\sigma,\sigma,\psi).
\end{aligned}
\]
Therefore, $\dim(A)=10$. However, we know that the algebra of logical operators should be $A_0=\text{Mat}_3$ which has $\dim(A_0)=9$, so $A$ is too large. This means that we need to reduce the redundancy in $A$ by modding out some relations. Conveniently, this redundancy reduction turns out to be equivalent to acting on $A$ by a projector $P$, which kills the subspace $(1-P)A$ and preserves its complement $PA$.

Before discussing where the relations come from, we want to first answer a question: How do we know whether we have found sufficiently many relations so that $PA$ is small enough? For a topological or fractonic order, its algebra of logical operators should be $\text{Mat}_n$ for some $n$. Conversely, a matrix algebra $\text{Mat}_n$ has the property that no more redundancy can be modded out (Definition~\ref{def:simple} and Lemma~\ref{lemma:mat_simple}). Therefore, the redundancy reduction stops if and only if $PA$ is a matrix algebra.

Furthermore, all of the algebras in the physical models in this paper have the additional property of being so-called semisimple.
\begin{definition}\label{def:ssimple}
    An algebra $A$ is \textit{semisimple} if it can be written as a direct sum
\begin{equation} \label{eq:ssimple_def}
    A=A_1\oplus\cdots\oplus A_m,
\end{equation}
    where each $A_i$ is a matrix algebra.
\end{definition}
The redundancy reduction amounts to finding a projector $P$ that kills all but one $A_i$, and then the true algebra of logical operators is this $A_i$. The kernel of $P$ consists of operators that we identify with 0, so we are taking a ``quotient'' of $A$ (see details in Appendix~\ref{app:details}).

Given $A$, its decomposition \eqref{eq:ssimple_def} can be performed systematically, but for the case of chiral Ising in this section, we will first write down the result:
\begin{equation} \label{eq:chiral_decomp}
    A=\text{Mat}_3\oplus\text{Mat}_1.
\end{equation}
A systematic derivation can be found in Section~\ref{sec:structure}. In this decomposition, we have
\[
\begin{aligned}
\text{Mat}_3&=\text{span}\{1+\psi_x,1+\psi_y,1+r,\\
&\!\qquad\qquad\sigma_x,\sigma_y,v(\sigma,\psi,\sigma),v(\psi,\sigma,\sigma),\\
&\!\qquad\qquad v(\sigma,\sigma,1),v(\sigma,\sigma,\psi)\},\\
\text{Mat}_1&=\text{span}\{1-r\},
\end{aligned}
\]
where
\begin{equation} \label{eq:r}
    r=\frac{1}{2}\left(1+\psi_x+\psi_y-\psi_x\psi_y\right).
\end{equation}
The 9 spanning elements of $\text{Mat}_3$ are not very important, but the element $r$ will be useful throughout this paper.

Given the decomposition \eqref{eq:chiral_decomp}, clearly we want to define the projector $P$ such that $PA=\text{Mat}_3$. However, if we didn't know that $A_0=\text{Mat}_3$ in the first place, then we would need to justify this choice of $P$. To do so, we note that $A$ is obtained only using fusion rules, $F$-symbols and $R$-symbols, while further information such as the topology of the torus hasn't been fully utilized. Indeed, we can put a contractible $\sigma$-loop ``around the corners'' of the torus, reduce it to a sum of elementary operators on the one hand, and demand it be equal to the quantum dimension $\sqrt{2}$ of $\sigma$ on the other hand. Using red lines for $\sigma$-strings and blue lines for $\psi$-strings, the reduction to elementary operators is performed as follows:
\begingroup
\allowdisplaybreaks
\begin{widetext}
\begin{equation}\label{eq:corner}
\begin{aligned}
    \frac{1}{\sqrt{2}}\vcenter{\hbox{
\begin{tikzpicture}
    \draw (-1, -1) rectangle (1, 1);
    \draw[red] (-0.75, -1) arc (0:90:0.25);
    \draw[red] (1, -0.75) arc (90:180:0.25);
    \draw[red] (0.75, 1) arc (180:270:0.25);
    \draw[red] (-1, 0.75) arc (270:360:0.25);
\end{tikzpicture}}}
\ &=\frac{1}{\sqrt{2}}\vcenter{\hbox{
\begin{tikzpicture}
    \draw (-1, -1) rectangle (1, 1);
    \draw[red] (-1, 0.1) -- (-0.35, 0.1);
    \draw[red] (-1, -0.1) -- (-0.35, -0.1);
    \draw[red] (1, 0.1) -- (0.35, 0.1);
    \draw[red] (1, -0.1) -- (0.35, -0.1);
    \draw[red] (-0.1, -1) -- (-0.1, -0.35);
    \draw[red] (0.1, -1) -- (0.1, -0.35);
    \draw[red] (-0.1, 1) -- (-0.1, 0.35);
    \draw[red] (0.1, 1) -- (0.1, 0.35);
    \draw[red] (-0.1, -0.35) arc (0:90:0.25);
    \draw[red] (0.35, -0.1) arc (90:180:0.25);
    \draw[red] (0.1, 0.35) arc (180:270:0.25);
    \draw[red] (-0.35, 0.1) arc (270:360:0.25);
\end{tikzpicture}}} \ =\frac{1}{\sqrt{2}}\sum_{a,b}\sqrt{\frac{d_a d_b}{d_\sigma^4}}
\vcenter{\hbox{
\begin{tikzpicture}
    \draw (-1, -1) rectangle (1, 1);
    \draw[red] (-0.1, -0.35) arc (0:90:0.25);
    \draw[red] (0.35, -0.1) arc (90:180:0.25);
    \draw[red] (0.1, 0.35) arc (180:270:0.25);
    \draw[red] (-0.35, 0.1) arc (270:360:0.25);
    \draw[red] (0.1, 0.35) arc (0:180:0.1);
    \draw[red] (-0.35, 0.1) arc (90:270:0.1);
    \draw[red] (-0.1, -0.35) arc (180:360:0.1);
    \draw[red] (0.35, -0.1) arc (-90:90:0.1);
    \draw (-1, 0) -- (-0.45, 0);
    \draw (0.45, 0) -- (1, 0);
    \draw (0, -1) -- (0, -0.45);
    \draw (0, 0.45) -- (0, 1);
    \node[above] at (-0.7, 0) {$a$};
    \node[right] at (0, -0.7) {$b$};
\end{tikzpicture}}}\\
&=\frac{1}{2\sqrt{2}}\left(
\vcenter{\hbox{
\begin{tikzpicture}
    \draw (-1, -1) rectangle (1, 1);
    \draw[red] (0.25, 0) arc (0:360:0.25);
\end{tikzpicture}}}
\ +
\vcenter{\hbox{
\begin{tikzpicture}
    \draw (-1, -1) rectangle (1, 1);
    \draw[red] (0.25, 0) arc (0:360:0.25);
    \draw[blue] (-1, 0) -- (-0.25, 0);
    \draw[blue] (0.25, 0) -- (1, 0);
\end{tikzpicture}}}
\ +
\vcenter{\hbox{
\begin{tikzpicture}
    \draw (-1, -1) rectangle (1, 1);
    \draw[red] (0.25, 0) arc (0:360:0.25);
    \draw[blue] (0, -1) -- (0, -0.25);
    \draw[blue] (0, 0.25) -- (0, 1);
\end{tikzpicture}}}
\ +
\vcenter{\hbox{
\begin{tikzpicture}
    \draw (-1, -1) rectangle (1, 1);
    \draw[red] (0.25, 0) arc (0:360:0.25);
    \draw[blue] (-1, 0) -- (-0.25, 0);
    \draw[blue] (0.25, 0) -- (1, 0);
    \draw[blue] (0, -1) -- (0, -0.25);
    \draw[blue] (0, 0.25) -- (0, 1);
\end{tikzpicture}}}
\ \right)\\
&=\frac{1}{2\sqrt{2}}\left(
\vcenter{\hbox{
\begin{tikzpicture}
    \draw (-1, -1) rectangle (1, 1);
    \draw[red] (0.25, 0) arc (0:360:0.25);
\end{tikzpicture}}}
\ +
\vcenter{\hbox{
\begin{tikzpicture}
    \draw (-1, -1) rectangle (1, 1);
    \draw[red] (0.25, 0) arc (0:360:0.25);
    \draw[blue] (-1, 0) -- (-0.25, 0);
    \draw[blue] (0.25, 0) -- (1, 0);
\end{tikzpicture}}}
\ +
\vcenter{\hbox{
\begin{tikzpicture}
    \draw (-1, -1) rectangle (1, 1);
    \draw[red] (0.25, 0) arc (0:360:0.25);
    \draw[blue] (0, -1) -- (0, -0.25);
    \draw[blue] (0, 0.25) -- (0, 1);
\end{tikzpicture}}}
\ +
\vcenter{\hbox{
\begin{tikzpicture}
    \draw (-1, -1) rectangle (1, 1);
    \draw[blue] (0, -1) arc (0:90:1);
    \draw[red] (0.25, 0) arc (0:360:0.25);
    \draw[blue] (0.25, 0) -- (1, 0);
    \draw[blue] (0, 0.25) -- (0, 1);
\end{tikzpicture}}}
\ \right)\\
&=\frac{1}{2\sqrt{2}}\left(\sqrt{2}+\sqrt{2}\psi_x+\sqrt{2}\psi_y+\sqrt{2}v(\psi,\psi,1)\right)\\
&=\frac{1}{2}(1+\psi_x+\psi_y-\psi_x\psi_y)\\
&=r,
\end{aligned}
\end{equation}
\end{widetext}
\endgroup
\noindent where we moved the $\sqrt{2}$ to the denominator. In this calculation, we first moved the $\sigma$-strings close together, and then fused the parallel $\sigma$-strings to obtain four outcomes (second line). The result is demanding $r=1$. In other words, we identify $1-r$ with 0 by taking the projector $P=(1+r)/2$, which precisely kills $1-r$. The same relation was also found in Ref.~\onlinecite{Aasen_2016}. We can repeat the same calculation for a 1-loop or a $\psi$-loop ``around the corners'', but in the end we obtain tautological relations. Only non-abelian anyons can give non-trivial relations.

To conclude this section, we summarize the operator algebra approach as follows:
\begin{protocol}\label{thm:summary}
    Suppose we have a topological or fractonic order.
\begin{enumerate}[nolistsep, leftmargin=*]
    \item We take a set of logical operators that span the space of all logical operators but are not necessarily linearly independent.
    
    \item We reduce the redundancy of these logical operators with $F$- and $R$-moves as much as possible. Then we take the formal algebra $A$ over the remaining operators, which is a semisimple algebra. In a $(2+1)$D topological order, if we take the operators $v(a,b,c)$ as in \eqref{eq:trivalent}, then these operators have no such redundancy and there is no need for this step.
    
    \item We find relations in $A$ by physical argument. In a $(2+1)$D topological order, the relations are given by loops of (non-abelian) anyons ``around the corners''. For Ising cage-net, we will see that the relations are given by cage structures of non-abelian strings. We then mod out the relations by acting with the corresponding projector $P$. If $PA$ is a matrix algebra, then the true algebra of logical operators is $A_0=PA$. In Section~\ref{sec:structure}, we will discuss a quick way to find $P$.
\end{enumerate}
\end{protocol}

\section{Structure of semisimple algebra}
\label{sec:structure}

The correctness of the decomposition \eqref{eq:chiral_decomp} can be checked by hand, but this is far from systematic. At the same time, we also do not have a systematic method for converting relations to projectors. In this section, we resolve these two issues by discussing the structure of a semisimple algebra, and provide an efficient way to compute projectors. Several statements in this section will be used in the calculations in later sections.

In the decomposition \eqref{eq:ssimple_def} of a semisimple algebra $A$, each component $A_i$ has its own multiplicative identity $P_i$, called a primitive central projector of $A$.
\begin{definition}
    An element $x\in A$ is \textit{central} if $[x,y]=0$ for all $y\in A$. The set of all central elements of $A$ is the \textit{center} of $A$, written as $Z(A)$. A central element $x\in Z(A)$ is a \textit{central projector} if $x^2=x$. A central projector $x$ is \textit{primitive} if for all central projector $y\in A$, we have $xy=0$ or $x$.
\end{definition}
The primitive central projectors $P_i$ have the property that every central projector $Q$ can be written as \[Q=\sum_i \lambda_i P_i,\] where $\lambda_i=0$ or 1. If we represent $A$ as block-diagonal matrices, then a central projector is the identity of several blocks, and a primitive one occupies exactly one block. It behaves like a projector in the usual sense when acting on $A$ by left multiplication (equivalent to right multiplication and conjugation since $Q$ is central).


In principle, given a basis $\{v_\alpha\}$ of $A$ and structure constants $f_{\alpha\beta}^\gamma$ defined by
\begin{equation}\label{eq:structure_const}
    v_\alpha v_\beta=\sum_\gamma f_{\alpha\beta}^\gamma v_\gamma,
\end{equation}
the central projectors are the solutions to the equations
\begin{equation}\label{eq:CI_solve}
    \begin{aligned}[c]
    [x,v_\alpha] &=0 \text{ for all }\alpha,\\
    x^2&=x.
    \end{aligned}
\end{equation}
If the solutions are $\{Q_k\}$, then the primitive ones form the subset $\{P_i\}\subset\{Q_k\}$ of maximal size such that $P_iP_j=0$ for all $i\neq j$. We then obtain the decomposition \eqref{eq:ssimple_def} where $A_i=P_iA$.

Next, we discuss the conversion of relations into projectors. In this paper, the relations obtained from physical argument happen to all be central in $A$. It also happens that a simply linear rescaling is enough to convert all the relations into central projectors, e.g.\ the rescaling of $1-r$ into $(1-r)/2$ for chiral Ising. Suppose that we have relations $Q_1,\ldots,Q_m$ where each $Q_k$ is a central projector. Then the overall projector is
\begin{equation}\label{eq:trick}
    P=(1-Q_1)\cdots (1-Q_m). 
\end{equation}
Such a projector can also be constructed without the assumption that $Q_k$ is central, and this construction is discussed in Appendix~\ref{app:details}.

Following this procedure, we find the primitive central projectors of chiral Ising to be \[P_1=\frac{1}{2}(1+r),\ P_2=\frac{1}{2}(1-r).\] Applying \eqref{eq:trick} with $Q=P_2$, we find \[PA=(1-P_2)A=P_1A=\text{Mat}_3,\] which is the matrix algebra we want.

In the remaining sections, \eqref{eq:trick} will be used constantly for computing projectors.

\section{Doubled Ising and condensation}
\label{sec:SN}

Ising cage-net is obtained via p-loop condensation, an example of Bose-Einstein condensation. In this section, we discuss the topic of condensation in the operator algebra approach by studying the simple example of condensation in doubled Ising. 

As explained in Section~\ref{sec:H}, doubled Ising is a stack of two copies of chiral Ising, whose anyons are 1, $\sigma$, $\psi$ and 1, $\bar\sigma$, $\bar\psi$, respectively. Now suppose we want to condense the boson $\psi\bar\psi$. For an anyon to remain deconfined upon condensation, it must braid trivially with $\psi\bar\psi$. Such anyons are $1=\psi\bar\psi$, $\psi=\bar\psi$ and $\sigma\bar\sigma$. Furthermore, $\sigma\bar\sigma$ is no longer a simple particle, but instead ``splits'' into two anyons $\sigma\bar\sigma=e+m$. To see why, note that $\sigma\bar\sigma$ is the fusion product of two Majorana modes and hence a (complex) fermion mode. The parity $p$ of this fermion mode can be 0 (unfilled) or 1 (filled), and braiding with $\sigma$ or $\bar\sigma$ switches the value of $p$, so $p$ is not a good quantum number in doubled Ising. However, if $\psi\bar\psi$ is condensed then both $\sigma$ and $\bar\sigma$ are confined, and therefore $p$ becomes a good quantum number that distinguishes the unfilled fermion mode (anyon $e$) from the filled (anyon $m$). The resulting topological order is the toric code\cite{Burnell_2012}.

It turns out that the operator algebra approach provides a nice description of condensation and, in particular, the ``splitting'' of anyons. To begin with, we follow Steps~1 and 2 of Protocol~\ref{thm:summary} to obtain a semisimple algebra $A$ with $\dim(A)=100$. Since doubled Ising is two copies of chiral Ising, we can find the decomposition of $A$ by taking a tensor product:
\begin{align}
    A&=\left(\text{Mat}_3\oplus\text{Mat}_1\right)^{\otimes 2} \nonumber\\
    &=\text{Mat}_9\oplus\text{Mat}_3\oplus\text{Mat}_3\oplus\text{Mat}_1. \label{eq:ssimple_doubled}
\end{align}
The quantum dimensions of $\sigma$ and $\bar\sigma$ give us two relations $r=1$ and $\bar r=1$, where
\begin{align*}
    r&=\frac{1}{2}\left(1+\psi_x+\psi_y-\psi_x\psi_y\right),\\
    \bar r&=\frac{1}{2}\left(1+\bar\psi_x+\bar\psi_y-\bar\psi_x\bar\psi_y\right).
\end{align*}
By \eqref{eq:trick}, these relations give rise to a projector \[P=\frac{1}{4}(1+r)(1+\bar r),\] and $PA=\text{Mat}_9$ is the correct (ground space) operator algebra of doubled Ising. Of course, $\sigma\bar\sigma$ is also a non-abelian anyon, and it gives a relation $r\bar r=1$, but we don't need to consider this relation separately because $r=1=\bar r$ already implies $r\bar r=1$.

To condense $\psi\bar\psi$, we want to identify $\psi_x\bar\psi_x$ and $\psi_y\bar\psi_y$ with 1 and understand the consequence of doing so. Let $M$ be the subalgebra of $A$ generated by $\psi_x\bar\psi_x$ and $\psi_y\bar\psi_y$. $M$ is an \textit{abelian} subalgebra since we have $[x,y]=0$ for all $x,y\in M$. Upon condensation, the logical operators that remain ``deconfined'' are those that commute with $M$. Such deconfined operators form the \textit{commutant} of $M$, which is a semisimple subalgebra of $A$ defined as \[M'=\{x\in A\,| \,[x,y]=0\ \text{for all}\ y\in M\}.\] Since $M$ is abelian, we have $M\subset M'$. To be precise, $M'$ is spanned by elementary operators $v(a,b,c)$ where $a$ and $b$ take values in $\{1,\psi,\bar\psi,\psi\bar\psi,\sigma\bar\sigma\}$. A straightforward calculation shows $\dim(M')=28$. By analyzing the primitive central projectors of $M'$ using \eqref{eq:CI_solve}, we can decompose $M'$ as
\begin{equation} \label{eq:ssimple_double_com}
    M'=\left(\text{Mat}_3\oplus 3\text{Mat}_2\right)\oplus 3\text{Mat}_1\oplus 3\text{Mat}_1\oplus\text{Mat}_1,
\end{equation}
where $3\text{Mat}_2$ means $\text{Mat}_2\oplus\text{Mat}_2\oplus\text{Mat}_2$, etc. Here, the summands are organized in correspondence with the summands in \eqref{eq:ssimple_doubled}, i.e.\ $\left(\text{Mat}_3\oplus 3\text{Mat}_2\right)$ is a subalgebra of the $\text{Mat}_9$ in \eqref{eq:ssimple_doubled}, the first $3\text{Mat}_1$ is a subalgebra of the first $\text{Mat}_3$ in \eqref{eq:ssimple_doubled}, and so on.

Next, we need to mod out all relations we know. Firstly, we have the quantum dimension of $\sigma\bar\sigma$, which demands $r\bar r=1$. By \eqref{eq:trick}, this gives a projector \[P_{12}=\frac{1}{2}(1+r\bar r).\] We chose the notation $P_{12}$ for consistency with the discussion in Section~\ref{sec:1f-ICN}. Now we note that \[P_{12}A=\text{Mat}_9\oplus\text{Mat}_1,\] since $r$ and $\bar r$ both act as $+1$ on $\text{Mat}_9$, and both act as $-1$ on $\text{Mat}_1$. Therefore, by restricting the action of $(1+r\bar r)/2$ to $M'$, we have \[P_{12}M'=\left(\text{Mat}_3\oplus 3\text{Mat}_2\right)\oplus\text{Mat}_1.\] Secondly, we have the condensation which demands $\psi_x\bar\psi_x=1$ and $\psi_y\bar\psi_y=1$. Again by \eqref{eq:trick}, these two relations give a projector \[P_\text{c}=\frac{1}{4}(1+\psi_x\bar\psi_x)(1+\psi_y\bar\psi_y),\] where the subscript ``c'' stands for ``condensation''. Thus the overall projector is \[P=P_\text{c}P_{12},\] and we need to check the action of $P_\text{c}$ on $\left(\text{Mat}_3\oplus 3\text{Mat}_2\right)$ and on $\text{Mat}_1$. The latter is straightforward: $\text{Mat}_1$ is spanned by $(1-r)(1-\bar r)$, and explicit calculation shows \[P_\text{c}(1-r)(1-\bar r)=(1-r)(1-\bar r).\] Therefore, $\text{Mat}_1$ is in $PM'$. On the other hand, let $Q_0=(1+r)(1+\bar r)/4$ be the central projector that projects $A$ onto $\text{Mat}_9$. Since both $P_\text{c}$ and $Q_0$ are central projectors, so is $P_\text{c}Q_0$, and we also claim that $P_\text{c}Q_0$ is primitive. This is a consequence of the following lemma:
\begin{lemma}\label{thm:center}
    Let $B$ be a matrix algebra, $N$ an abelian subalgebra of $B$, and $N'$ the commutant of $N$. Then we have $Z(N')=N$.
\end{lemma}
It is easy to see that $N\subset Z(N')$, and Lemma~\ref{thm:center} says that the two are actually equal. Technically, $N$ must satisfy another condition, and Lemma~\ref{thm:center2} in Appendix~\ref{app:details} explains this point more rigorously. Applying Lemma~\ref{thm:center} with $B=Q_0A$ and $N=Q_0M$, we know that $Z(Q_0M')$ is generated by $\psi_x\bar\psi_xQ_0$ and $\psi_y\bar\psi_yQ_0$. It is then straightforward to use the prescription in Section~\ref{sec:structure} to find the primitive projectors from the central elements, and indeed $P_\text{c}Q_0$ is one of them. Thus we know that $P_\text{c}Q_0M'$ is a matrix algebra, and we need to determine whether it is $\text{Mat}_3$ or one of the three copies of $\text{Mat}_2$. For this purpose, we note that for any operator $x\in A$, we can represent $xQ_0$ as a $9\times9$ matrix $\rho_9(xQ_0)$, or $\rho_9(x)$ for short. The subscript $l$ in $\rho_l$ indicates the matrix dimension. A systematic way to determine this representation $\rho_9$ can be found in Appendix~\ref{app:matrix}, but here we will start with a $3\times3$ matrix representation $\rho_3$ of operators in chiral Ising:
\begin{equation}\label{eq:matrix3}
\begin{aligned}
\rho_3(\psi_x)&=
\begin{pmatrix}
1 & &\\
& 1 &\\
& & -1
\end{pmatrix},\ \rho_3(\sigma_x)=
\begin{pmatrix}
0 & \sqrt{2} & 0\\
\sqrt{2} & 0 & 0\\
0 & 0 & 0
\end{pmatrix},\\
\rho_3(\psi_y)&=
\begin{pmatrix}
1 & &\\
& -1 &\\
& & 1
\end{pmatrix},\ \rho_3(\sigma_y)=
\begin{pmatrix}
0 & 0 & \sqrt{2}\\
0 & 0 & 0\\
\sqrt{2} & 0 & 0
\end{pmatrix}.
\end{aligned}
\end{equation}
The correctness of this representation can be confirmed by hand or by following the discussion in Appendix~\ref{app:matrix}. The operators in $\text{Mat}_9$ can be obtained by tensoring the above matrices. In particular, $\rho_9(Q_0)$ is the $9\times 9$ identity, and we find $\rho_9(P_\text{c}Q_0)$ to be a diagonal matrix
\begin{equation}\label{eq:condense_DI}
    \rho_9(P_\text{c}Q_0)=\text{diag}(1,0,0,0,1,0,0,0,1).
\end{equation}
Since $\text{tr}(\rho_9(P_\text{c}Q_0))=3$, we have $P_\text{c}Q_0M'=\text{Mat}_3$. To summarize, we have
\begin{equation}\label{eq:pre_split_DI}
    PM'=\text{Mat}_3\oplus\text{Mat}_1,
\end{equation}
where the projector $P$ accounts for deconfined anyons and the condensation condition.

The bottom line of \eqref{eq:pre_split_DI} is that even after modding out all the relations we know, we still do not obtain a matrix algebra. However, we know that the algebra of logical operators must be a matrix algebra, so we need to do something to $PM'$. For this purpose, we visualize $PM'$ as block-diagonal matrices embedded in $\text{Mat}_4$:
\begin{equation}\label{eq:fill_blank}
    PM'=\vcenter{\hbox{
    \begin{tikzpicture}
        \draw (-1, -1) rectangle (1, 1);
        \draw[pattern=north east lines] (-1, -0.5) rectangle (0.5, 1);
        \draw[pattern=north east lines] (0.5, -1) rectangle (1, -0.5);
    \end{tikzpicture}}}~.
\end{equation}
We have the following observation: The splitting of $\sigma\bar\sigma$ precisely ``fills the blanks'' in \eqref{eq:fill_blank} to turn $\text{Mat}_3\oplus\text{Mat}_1$ into $\text{Mat}_4$.

To justify this observation, we will work out a $4\times4$ matrix representation $\rho_4$ of, say, $e_x$ and compare it with the known result from the toric code. By \eqref{eq:condense_DI}, the $\text{Mat}_3$ block of an element $x\in PM'$ is obtained by taking rows and columns 1, 5 and 9 from $\rho_9(x)$. On the other hand, the $\text{Mat}_1$ block of $x\in PM'$ is determined by its action on the generator $(1-r)(1-\bar r)$ of $\text{Mat}_1$. For example,
\begin{align*}
    \psi_x(1-r)(1-\bar r)&=-(1-r)(1-\bar r),\\
    \sigma_x\bar\sigma_x(1-r)(1-\bar r)&=0.
\end{align*}
Using this method, we find the $\rho_4$ representations of some operators in $PM'$ to be
\begin{align*}
    \rho_4(\psi_x)&=
    \begin{pmatrix}
    1 & & &\\
    & 1 & &\\
    & & -1 &\\
    & & & -1
    \end{pmatrix},\ \rho_4(\sigma_x\bar\sigma_x)=
    \begin{pmatrix}
    0 & 2 & 0 & 0\\
    2 & 0 & 0 & 0\\
    0 & 0 & 0 & 0\\
    0 & 0 & 0 & 0
    \end{pmatrix},\\
    \rho_4(\psi_y)&=
    \begin{pmatrix}
    1 & & &\\
    & -1 & &\\
    & & 1 &\\
    & & & -1
    \end{pmatrix},\ \rho_4(\sigma_y\bar\sigma_y)=
    \begin{pmatrix}
    0 & 0 & 2 & 0\\
    0 & 0 & 0 & 0\\
    2 & 0 & 0 & 0\\
    0 & 0 & 0 & 0
    \end{pmatrix}.
\end{align*}
We want to use physical argument to find $\rho_4(e_x)$. We have equations
\begin{align*}
    \rho_4(e_x)^\dagger&=\rho_4(e_x),\\
    (1+\rho_4(\psi_x))\rho_4(e_x)&=\rho_4(\sigma_x\bar\sigma_x),\\
    \rho_4(e_x)\rho_4(\psi_y)&=-\rho_4(\psi_y)\rho_4(e_x),\\
    \rho_4(e_x)^2&=1.
\end{align*}
Line~1 says that $e$ is its own antiparticle; line~2 comes from $\psi\times e=m$ and $\sigma\bar\sigma=e+m$; line~3 says that $e$ and $\psi$ braid with a $-1$ phase; line~4 comes from the fusion rule of $e$. The most general solution is
\[
\rho_4(e_x)=
\begin{pmatrix}
0 & 1 & &\\
1 & 0 & &\\
& & 0 & e^{i\theta}\\
& & e^{-i\theta} & 0
\end{pmatrix}.
\]
As expected, $\rho_4(e_x)$ has entries $e^{\pm i\theta}$ in the ``blank'' areas of \eqref{eq:fill_blank}. There is no way to fix $\theta$, since conjugation by
\[
U=
\begin{pmatrix}
1 & & &\\
& 1 & &\\
& & 1 &\\
& & & e^{i\phi}
\end{pmatrix}
\]
acts trivially on $\text{Mat}_3\oplus\text{Mat}_1$ but non-trivially on $\text{Mat}_4$, mapping $\theta$ to $\theta\pm\phi$. Without loss of generality, we choose $\theta=0$. This gives
\[
\rho_4(e_x)=
\begin{pmatrix}
0 & 1 & &\\
1 & 0 & &\\
& & 0 & 1\\
& & 1 & 0
\end{pmatrix},\ 
\rho_4(m_x)=
\begin{pmatrix}
0 & 1 & &\\
1 & 0 & &\\
& & 0 & -1\\
& & -1 & 0
\end{pmatrix}.
\]
Using the same method while demanding $\rho_4(e_y)$ commute with $\rho_4(e_x)$, we find
\[
\rho_4(e_y)=
\begin{pmatrix}
0 & 0 & 1 & 0\\
0 & 0 & 0 & 1\\
1 & 0 & 0 & 0\\
0 & 1 & 0 & 0
\end{pmatrix},\ 
\rho_4(m_y)=
\begin{pmatrix}
0 & 0 & 1 & 0\\
0 & 0 & 0 & -1\\
1 & 0 & 0 & 0\\
0 & -1 & 0 & 0
\end{pmatrix}.
\]
One may check that these indeed obey the (ground space) operator algebra of the toric code. Moreover, they generate matrices such as
\[
\begin{pmatrix}
0 & 0 & 0 & 1\\
0 & 0 & 0 & 0\\
0 & 0 & 0 & 0\\
0 & 0 & 0 & 0
\end{pmatrix}
=\frac{1}{4}\rho_4(\sigma_x\bar\sigma_x)[\rho_4(e_y)-\rho_4(m_y)],
\]
and hence all other matrices with entries in the ``blank'' areas of \eqref{eq:fill_blank}.

To conclude this section, we summarize condensation in the operator algebra approach as follows:
\begin{protocol}\label{thm:conjecture}
    Let $A$ be the semisimple algebra of a topological or fractonic order, and suppose that $\{a\}$ is a set of bosons to be condensed.
\begin{enumerate}[nolistsep, leftmargin=*]
    \item We define $M$ as the subalgebra of logical operators of $\{a\}$. If $\{a\}$ can be condensed simultaneously, then $M$ is always abelian.
    
    \item Let $M'$ be the commutant of $M$. We define $P$ as the projector due to the condensation condition, relations due to deconfined anyons as well as relations from other physical arguments. For Ising cage-net, the physical arguments will come from the plaquette and cage terms of the Hamiltonian \eqref{eq:H}. We then take the algebra $PM'$.
    
    \item If the semisimple algebra \[PM'=\text{Mat}_{d_1}\oplus\cdots\oplus\text{Mat}_{d_m}\] has more than one component, then certain operators must split. The result of splitting is a matrix algebra \[A_0=\text{Mat}_{d_1+\cdots+d_m},\] which is obtained by ``filling the blanks'' in the matrix representation of $PM'$. The correctness of this operation can be confirmed manually for all the $(2+1)$D models in this paper, and we conjecture that this holds for all topological or fractonic orders.
\end{enumerate}
\end{protocol}

\section{GSD of One-foliated Ising cage-net}
\label{sec:1f-ICN}

We will discuss one more $(2+1)$D topological order in this section before going to Ising cage-net in Section~\ref{sec:ICN}. In particular, we will show a method of computing the GSD using a Cartan subalgebra, which is very convenient when applied to Ising cage-net.

The model we want to discuss is called the one-foliated Ising cage-net model (``1-F Ising'' for short), which is constructed as follows: We take a stack of $2L$ copies of chiral Ising, and condense the boson $\Psi=\psi(1)\times\cdots\times\psi(2L)$, where $\psi(k)$ is the $\psi$ particle from the $k$th layer. The chirality of these copies of chiral Ising does not affect the GSD. The condensation of doubled Ising into the toric code in Section~\ref{sec:SN} is a special case of this construction with $L=1$.

\begin{figure}[t]
    \centering
    \includegraphics[scale = 0.1]{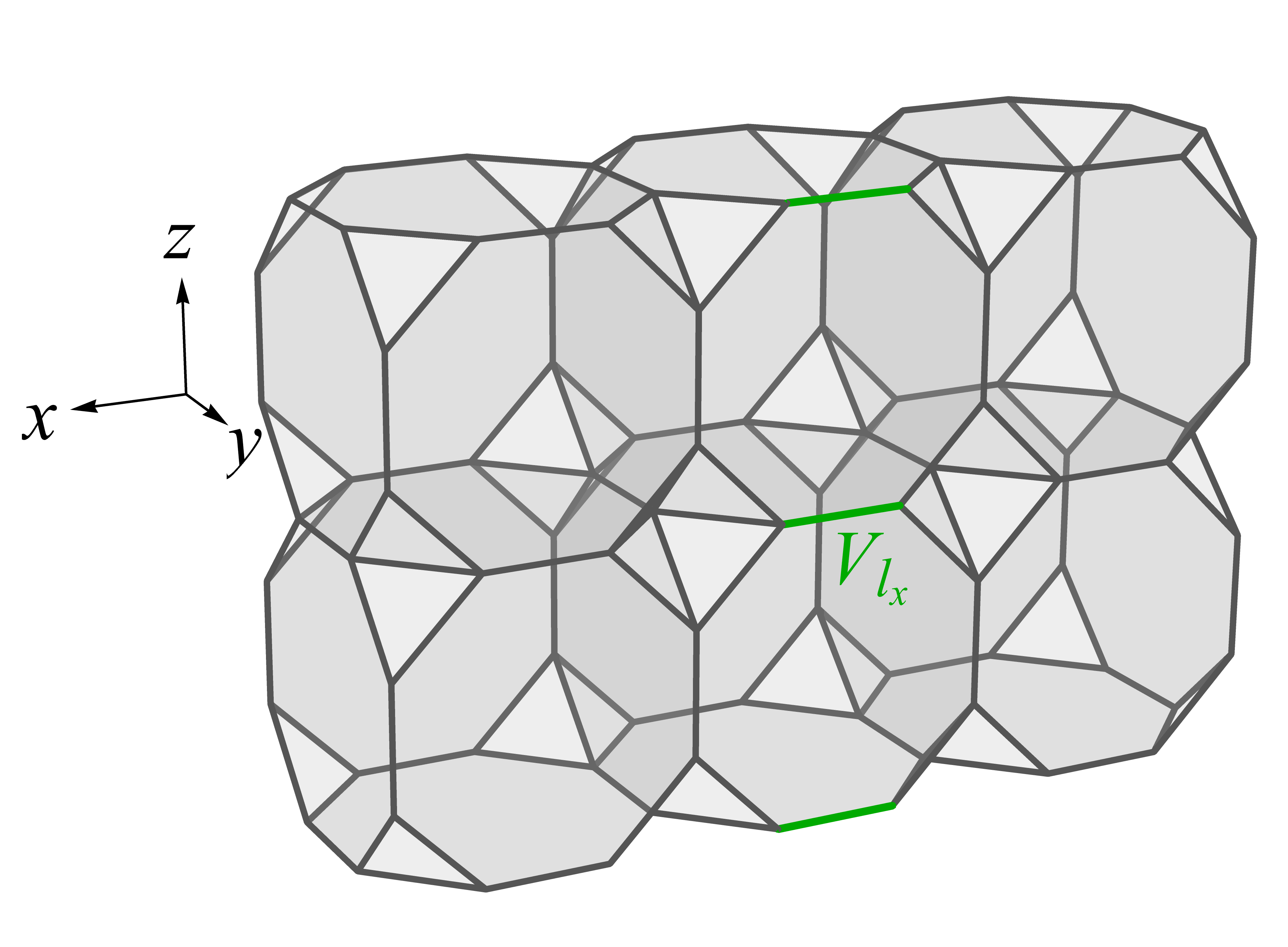}
    \captionsetup{justification=Justified}
    \caption{1-F Ising obtained from layers of doubled Ising on a square-octagon lattice, similar to the construction of Ising cage-net. Each plane is a layer of doubled Ising. The product of $V_{l_x}$ on the green edges (the set $S_x$ in \eqref{eq:1f_creation}) creates a pair of $\psi\bar\psi$ in each $xy$ plane. If we condense the $\Psi$ particles created this way, then we obtain 1-F Ising together with decoupled layers of doubled Ising in the $yz$ and $zx$ planes.}
    \label{fig:condense_1f}
\end{figure}

In the limit $L\to\infty$, 1-F Ising can be viewed as a fracton model, whose partially mobile excitations are planons. It is related to Ising cage-net as follows: In Ising cage-net, let $S_x$ be a set of principal edges $l_x$ related to each other by translation in the $z$ direction (green edges in Fig.~\ref{fig:condense_1f}). Then the operator
\begin{equation} \label{eq:1f_creation}
    \prod_{l_x\in S_x}V_{l_x},
\end{equation}
where $V_{l_x}$ is a condensation operator defined in \eqref{eq:condense_op}, creates a pair of $\psi\bar\psi$ anyons in each $xy$ plane. Therefore, $\Psi$ in the $xy$ plane is part of the condensate in Ising cage-net. So are $\Psi$ in the $yz$ and $zx$ planes. In this sense, 1-F Ising is the ``one-foliated'' version of Ising cage-net, while Ising cage-net is ``three-foliated''.

\subsection{GSD from anyon counting}

Since the 1-F Ising is a $(2+1)$D topological order, its GSD can be obtained by counting anyons. In order for an anyon to be deconfined upon condensation, it must contain an even number of $\sigma's$ in order to braid trivially with \(\Psi = \psi_1\times\cdots\times\psi_{2L}\). The only such particles are ones where there are an even number of layers with \(\sigma\). Additionally, we can attach \(\psi\) to any layer where there is no \(\sigma\), since that does not affect the braiding with \(\Psi\). The condensation of $\Psi$ identifies some pairs of anyons with each other, which reduces the number of distinct anyons. Finally, the anyon $\Sigma=\sigma(1)\times\cdots\times\sigma(2L)$ splits into two simple particles $\Sigma=e+m$ since the overall fermion parity of $\Sigma$ is a good quantum number. Another way to see this is to note that \(\Sigma \times \Sigma = 1 + \Psi + \cdots\), so the presence of two identity channels implies that \(\Sigma\) splits into two particles. These conditions give constraints on a label $a(1)\times\cdots\times a(2L)$ of an anyon.

We now count the number of such labelings. If we choose \(2k\) layers \(i_1,\ldots,i_{2k}\) to attach \(\sigma\) to, where $k=0,\ldots,L-1$, then there are \((2L-2k)\) places left to attach \(\psi\). It would seem, therefore, that there are \(2^{2L-2k}\) inequivalent ways to attach \(\psi\) to the layers. However, \(\Psi = 1\) reduces the number of distinct labelings by a factor of 2. For example, consider the case where $L = 4$. Here, the particles $\sigma_1 \sigma_2 \psi_3$ and $\sigma_1 \sigma_2 \psi_4$ are equivalent, since $\sigma_1 \sigma_2 \psi_3 = \sigma_1 \sigma_2 \times \psi_1 \psi_2 \psi_3 = \sigma_1 \sigma_2 \times \psi_4$ using $\Psi = \psi_1\psi_2\psi_3\psi_4 = 1$. This generalizes straightforwardly to an arbitrary number of layers and $\sigma$'s, halving the number of anyons in the theory. Therefore, there are \(\binom{2L}{2k} 2^{2L-2k-1}\) inequivalent ways to choose \(2k\) layers to place \(\sigma\) and attach 1's or \(\psi\)'s to the remaining layers. The case where $k=L$ needs to be considered separately. In this case, the anyon of interest is $\Sigma$, which splits into $e$ and $m$. Thus the total number of anyons in the theory (equal to the GSD) is
\begin{align*}
    \text{GSD}&=\sum_{k=0}^{L-1}{2L \choose 2k}2^{2L-2k-1}+2 \\
    &=\sum_{k=0}^{L}{2L \choose 2k}2^{2L-2k-1}+2-\frac{1}{2},
\end{align*}
where the \(+2\) accounts for the \(k=L\) case. To evaluate this, we use the binomial theorem
\begin{align*}
    (1+x)^{2L}+(1-x)^{2L}&= \sum_{j=0}^{2L} \binom{2L}{j} \left(x^{2L-j}+(-x)^{2L-j} \right) \\
    &=2\sum_{k=0}^{L} \binom{2L}{2k}x^{2L-2k}.
\end{align*}
This can be used to find the GSD:
\begin{align}
    \text{GSD} &=\sum_{k=0}^{L}{2L \choose 2k}2^{2L-2k-1}+2-\frac{1}{2} \nonumber \\
    &= \frac{1}{4}(1+2)^{2L}+\frac{1}{4}(1-2)^{2L}+\frac{3}{2} \nonumber\\
    &= \frac{1}{4}\left(9^L+7\right). \label{eq:1f_GSD}
\end{align}
The same result was obtained by K.\ Slagle, D.\ Aasen, D.\ Williamson and W.\ Shirley\cite{anyon_counting}.

\subsection{GSD from Cartan subalgebra}
\label{sec:1f_Cartan}

We now try to reproduce \eqref{eq:1f_GSD} using the operator algebra approach. Protocol~\ref{thm:conjecture} is based on the full algebra of 1-F Ising, but we delay this calculation to Section~\ref{sec:1f_full}. Instead, here we will compute the GSD using a so-called Cartan subalgebra of the full algebra.
\begin{definition}\label{def:Cartan}
    A subalgebra $C$ of an algebra $A$ is a \textit{Cartan subalgebra} if it is abelian and maximal. Abelian means that $[x,y]=0$ for all $x,y\in C$; maximal means that if any subalgebra $C'\subset A$ is abelian and $C\subset C'$, then $C'=C$.
\end{definition}
Note that this definition is not entirely rigorous mathematically: There is another condition on $C$ which we did not mention, and this condition holds for the choice of $C$ that we will use later. Definition~\ref{def:Cartan2} in Appendix~\ref{app:details} explains this extra condition.

A Cartan subalgebra is related to the GSD by the following lemma:
\begin{lemma}\label{thm:Cartan}
    Let $A_0$ be a matrix algebra, $C_0\subset A_0$ a Cartan subalgebra. Then $\dim(C_0)^2=\dim(A_0)$. In particular, if $A_0$ is a (ground space) operator algebra, then $\text{GSD}=\dim(C_0)$.
\end{lemma}
To understand this lemma with an example, take $C_0$ to be the set of diagonal matrices in $A_0$. The lemma is obvious in this case.

For $2L$ copies of chiral Ising with semisimple algebra \[A=\left(\text{Mat}_3\oplus\text{Mat}_1\right)^{\otimes 2L},\] we have a convenient choice of a Cartan subalgebra $C$, which is spanned by the elementary operators with no $\sigma$. To compute the GSD, we want to understand the transition from $C$ to $C_0$. Our approach will be similar to Steps~1 and 2 of Protocol~\ref{thm:conjecture}, although we will adapt these steps to fit with the Cartan subalgebra. Let $M$ be the subalgebra of $A$ generated by $\Psi_x$ and $\Psi_y$ (the condensate). In the commutant $M'$ of $M$, we have central projectors
\begin{equation}\label{eq:condense_1f}
    P_\text{c}=\frac{1}{4}(1+\Psi_x)(1+\Psi_y)
\end{equation}
due to condensation, and
\begin{equation}\label{eq:pair_1f}
    P_{ij}=\frac{1}{2}(1+r(i)r(j))
\end{equation}
due to deconfined anyons $\sigma(i)\sigma(j)$, where $1\le i<j\le 2L$ and \[r(i)=\frac{1}{2}\left(1+\psi_x(i)+\psi_y(i)-\psi_x(i)\psi_y(i)\right).\] Although there are also non-abelian anyons consisting of more than two $\sigma$'s and possibly $\psi$'s, for the purpose of constructing projectors it suffices to only consider pairs of $\sigma$'s. This is because e.g.\ $\sigma(1)\sigma(2)\sigma(3)\sigma(4)$ gives a relation $r(1)r(2)r(3)r(4)=1$, but this is already implied by $r(1)r(2)=1$ and $r(3)r(4)=1$.

From this point on, we will focus only on the Cartan subalgebra. Importantly, in this specific case we have $C\subset M'$, and the central projectors $P_\text{c}$ and $P_{ij}$ all map $C$ to $C$ since they also contain no $\sigma$. Meanwhile, we can argue physically that splitting does not enlarge the Cartan subalgebra. This is because the braiding of $\Sigma$ with e.g.\ $\psi(1)$ gives a $-1$ phase and thus the same holds for the anyons $e$ and $m$ split from $\Sigma$. Therefore, the entirety of $C_0$ can be obtained by projection on $C$. In other words, we have $C_0=PC$, where
\begin{equation}\label{eq:proj_1f}
    P=P_\text{c}\prod_{i<j}P_{ij}.
\end{equation}
Since $P$ is a projector, we have \[\text{GSD}=\dim(PC)=\text{tr}(P).\] We stress that the underlying vector space here is $C$, and that we are taking the trace of the action of $P$ on $C$.

To find $\text{tr}(P)$, we note that in principle, we can write
\begin{equation}\label{eq:proj_expansion}
    P=\sum_{a,b}\mu(a,b)v(a,b,a\times b),
\end{equation}
where neither $a$ nor $b$ contains any $\sigma$ (and hence $a\times b$ is unique), and $\mu(a,b)$ are some coefficients. We then observe that when $v(c,d,c\times d)\in C$ is multiplied by $v(a,b,a\times b)$ in \eqref{eq:proj_expansion}, the result \[\pm v(a\times c,b\times d,a\times b\times c\times d)\] is never proportional to $v(c,d,c\times d)$ unless $a=b=1$. As a result, only $v(1,1,1)$ contributes to $\text{tr}(P)$, so we only need to find the coefficient $\mu(1,1)$. For this purpose, we need to expand \eqref{eq:proj_1f}. Firstly, we use $r(i)^2=1$ to obtain
\begin{align*}
    \prod_{i<j}P_{ij}&=\frac{1}{2^{2L-1}}\sum_{k=0}^L \prod_{i_1<\cdots<i_{2k}}r(i_1)\cdots r(i_{2k})\\
    &=\frac{1}{2^{2L}}\left[\prod_{i=1}^{2L}(1+r(i))+\prod_{i=1}^{2L}(1-r(i))\right].
\end{align*}
The first line is a sum of all products of an even number of $r(i)$'s; the second line can be interpreted as forcing the $r(i)$'s to be all $+1$ or all $-1$, which is a consequence of forcing each pair of the $r(i)$'s to be both $+1$ or both $-1$ due to $\{P_{ij}\}$. Thus
\begin{align*}
    P=\frac{1}{4}&(1+\Psi_x+\Psi_y+\Psi_x\Psi_y)\\
    &\times\frac{1}{2^{2L}}\left[\prod_{i=1}^{2L}(1+r(i))+\prod_{i=1}^{2L}(1-r(i))\right].
\end{align*}
Now since
\begin{align*}
    1+r(i)=\frac{3}{2}+\frac{1}{2}\psi_x(i)+\frac{1}{2}\psi_y(i)-\frac{1}{2}\psi_x(i)\psi_y(i),\\
    1-r(i)=\frac{1}{2}-\frac{1}{2}\psi_x(i)-\frac{1}{2}\psi_y(i)+\frac{1}{2}\psi_x(i)\psi_y(i),
\end{align*}
the only four terms in the expansion of $\prod_i(1+r(i))$ that combines with one of 1, $\Psi_x$, $\Psi_y$ and $\Psi_x\Psi_y$ to contribute to $\mu(1,1)$ are
\begin{align*}
    \left(\frac{3}{2}\right)^{2L},\ \prod_i\left(\frac{1}{2}\psi_x(i)\right),&\ \prod_i\left(\frac{1}{2}\psi_y(i)\right),\\
    \text{and }&\prod_i\left(-\frac{1}{2}\psi_x(i)\psi_y(i)\right).
\end{align*}
Summing these up, we find that the contribution of the $\prod_i(1+r(i))$ part to $\mu(1,1)$ is $\left(9^L+3\right)/2^{4L+2}$. Similarly, the contribution of the $\prod_i(1-r(i))$ part is $4/2^{4L+2}$. Combining these together, we obtain \[\text{GSD}=\text{tr}(P)=2^{4L}\mu(1,1)=\frac{1}{4}\left(9^L+7\right),\] where we used the fact that $\dim(C)=2^{4L}$.

This calculation is almost entirely combinatorial and straightforward. However, it is also highly specific to simple examples such as Ising -- it relies on a nice Cartan subalgebra which is fixed by the central projectors and cannot be enlarged by splitting due to physical arguments.

\subsection{GSD from full algebra}
\label{sec:1f_full}

Finally for the discussion of 1-F Ising, we compute its GSD using Protocol~\ref{thm:conjecture}.

Again, we take $M$ to be the subalgebra of $A$ generated by $\Psi_x$ and $\Psi_y$, and $M'$ the commutant of $M$. We want to find $PM'$ where $P$ is given by \eqref{eq:proj_1f}. We will not try to decompose $M'$ into matrix algebras like we did for doubled Ising in Section~\ref{sec:SN}, since most of the components of $A$ will be killed by the central projectors $P_{ij}$, just like the two copies of $\text{Mat}_3$ in \eqref{eq:ssimple_doubled}. Instead, we will first discuss the action of $P_{ij}$ on $A$, and then apply $P_\text{c}$. To start with, we have \[\prod_{i<j}P_{ij}A=B_0\oplus B_1,\] where $B_0=\text{Mat}_{9^L}$ and $B_1=\text{Mat}_1$. This is because $\{P_{ij}\}$ forces the $r(i)$'s to be all $+1$ or all $-1$. As a result, we have \[\prod_{i<j}P_{ij}M'\subset B_0\oplus B_1,\] so we need to find the action of $P_\text{c}$ on $(B_0\oplus B_1)\cap M'$.

For the $B_1\cap M'$ part, we know that $B_1\subset Z(A)$ since $B_1$ is a $1\times 1$ block. Thus $B_1\cap M'=B_1$. Each of $\psi_x(i)$ and $\psi_y(i)$ acts on $B_1$ as $-1$, so $P_\text{c}$ preserves $B_1$. We conclude that $P_\text{c}(B_1\cap M')=\text{Mat}_1$.

For the $B_0\cap M'$ part, we will repeat what we did in Section~\ref{sec:SN} for $\text{Mat}_3\oplus 3\text{Mat}_2$, and use a matrix representation of $P_\text{c}$ to determine its action. Let \[Q_0=\frac{1}{2^{2L}}\prod_i (1+r(i))\] be the central projector that projects onto $B_0$. By Lemma~\ref{thm:center}, the central projector $P_\text{c}Q_0$ is primitive, and hence the algebra $P_\text{c}Q_0M'$ is a matrix algebra. On $B_0$, the action of operators such as $P_\text{c}$ has representation $\rho_{9^L}$. Thus we have $P_\text{c}Q_0M'=\text{Mat}_n$ where $n=\text{tr}(\rho_{9^L}(P_\text{c}Q_0))$. To find $n$, we use
\begin{align*}
    n&=\dim(\text{eigenspace }\rho_{9^L}(P_\text{c})=1)\\
    &=\dim(\text{eigenspace }\rho_{9^L}(\Psi_x)=\rho_{9^L}(\Psi_y)=+1).
\end{align*}
Let $D^{st}_{2L}$, where $s$, $t$ can be $+$ or $-$, be the dimension of the common eigenspace $\{w\}$ of $\rho_{9^L}(\Psi_x)$ and $\rho_{9^L}(\Psi_y)$ where $\rho_{9^L}(\Psi_x)w=sw$ and $\rho_{9^L}(\Psi_y)=tw$ (i.e.\ $\pm w$). From the representation $\rho_3$ of $\psi_x$ and $\psi_y$ in \eqref{eq:matrix3}, we find
\begin{equation}\label{eq:common_eigen}
    \begin{aligned}
    D^{++}_{2L}&=\frac{1}{4}\left(9^L+3\right),\\
    D^{+-}_{2L}&=D^{-+}_{2L}=D^{--}_{2L}=\frac{1}{4}\left(9^L-1\right).
    \end{aligned}
\end{equation}
We will show the calculation of $D^{+-}_{2L}$ as an example. Let $\{u_1,u_2,u_3\}$ be the standard basis for $\mathbb{C}^3$, and \[w=u_1^{\otimes k_1}\otimes u_2^{\otimes k_2}\otimes u_3^{\otimes k_3}.\] In order for $\rho_{9^L}(\Psi_x)w=+w$ and $\rho_{9^L}(\Psi_y)w=-w$, according to \eqref{eq:matrix3}, we must have $k_3$ odd, $k_2$ even, and hence $k_1$ odd. The number of such combinations of $(k_1,k_2,k_3)$ satisfying $k_1+k_2+k_3=2L$ can be found using the multinomial theorem:
\begin{align*}
    D^{+-}_{2L}&=\frac{1}{4}\left[(1+1+1)^{2L}-(1+1-1)^{2L}\right.\\
    &\qquad\ \left.+(1-1+1)^{2L}-(1-1-1)^{2L}\right]\\
    &=\frac{1}{4}\left(9^L-1\right).
\end{align*}
Using \eqref{eq:common_eigen}, we find $\text{tr}(\rho_{9^L}(P_\text{c}Q_0))=D^{++}_{2L}=\left(9^L+3\right)/4$. Although here we only made use of $D^{++}_{2L}$, the other $D$'s will be used in Section~\ref{sec:ICN_full}. 

Putting the $B_0\cap M'$ and $B_1\cap M'$ parts together, we conclude that \[PM'=\text{Mat}_{\left(9^L+3\right)/4}\oplus\text{Mat}_1.\] This is a semisimple algebra. Similar to what we did in Section~\ref{sec:SN} for condensation in doubled Ising, we can also find matrix representations of $e_x$, $m_x$, $e_y$ and $m_y$ and confirm that they have non-zero entries in the ``blank'' areas of $PM'$, but we omit this calculation here. The semisimple algebra then turns into a matrix algebra
\[A_0=\text{Mat}_{\left(9^L+7\right)/4},\] and $\text{GSD}=\left(9^L+7\right)/4$ as expected.

\section{GSD of Ising cage-net}
\label{sec:ICN}

In this section, we compute the GSD of Ising cage-net, first using a Cartan subalgebra, and then using the full algebra.

\begin{figure}[t]
    \centering
    \begin{tikzpicture}
        \draw (-0.3536, -1.3536) -- (0.3536, -0.6464) -- (0.3536, 1.3536) -- (-0.3536, 0.6464) -- cycle;
        \draw (-3, -1) rectangle (-1, 1);
        \draw (0.6464, -0.3536) -- (2.6464, -0.3536) -- (3.3536, 0.3536) -- (1.3536, 0.3536) -- cycle;
        \begin{axis}[hide axis, anchor=origin, disabledatascaling, at={(0pt,0pt)}, x=1cm,y=1cm]
            \addplot [domain=0:90] ({-0.3536*cos(x)+0.3536}, {-0.3536*cos(x)+sin(x)-0.6464});
            \addplot [domain=180:270] ({-0.3536*cos(x)-0.3536}, {-0.3536*cos(x)+sin(x)+0.6464});
            \addplot [domain=180:270] ({-0.3536*cos(x)+sin(x)+2.6464}, {-0.3536*cos(x)-0.3536});
            \addplot [domain=0:90] ({-0.3536*cos(x)+sin(x)+1.3536}, {-0.3536*cos(x)+0.3536});
        \end{axis}
        \draw (-2, -1) arc (0:90:1);
        \draw (-2, 1) arc (180:270:1);
        \draw (1.8779, 0.1221) -- (2.1221, -0.1221);
        \draw (-0.1221, 0.1221) -- (0.1221, -0.1221);
        \draw (-2.2929, -0.2929) -- (-1.7071, 0.2929);
        \node at (-0.05, -0.2) {$c^y$};
        \node at (2.2, 0.1) {$c^z$};
        \node at (-0.15, -0.85) {$a^y_z$};
        \node at (0.2, 0.5) {$b^y_x$};
        \node at (1.5, -0.18) {$a^z_x$};
        \node at (2.9, 0.18) {$b^z_y$};
        \node at (-1.9, -0.1) {$c^x$};
        \node at (-2.75, 0.18) {$a^x_y$};
        \node at (-2.18, 0.75) {$b^x_z$};
        
        \draw[->, -latex] (-4, 0) -- (-3.5, 0);
        \draw[->, -latex] (-4, 0) -- (-4, 0.5);
        \draw[->, -latex] (-4, 0) -- (-4.1768, -0.1768);
        \node at (-4.3, -0.3) {$x$};
        \node at (-3.3, 0) {$y$};
        \node at (-4, 0.7) {$z$};
    \end{tikzpicture}
    \captionsetup{justification=Justified}
    \caption{Constituents $v^x(a^x_y,b^x_z,c^x)$, $v^y(a^y_z,b^y_x,c^y)$ and $v^z(a^z_x,b^z_y,c^z)$ of an elementary operator in Ising cage-net. Arrows are not drawn since in Ising cage-net, every particle is its own antiparticle.}
    \label{fig:op3d}
\end{figure}
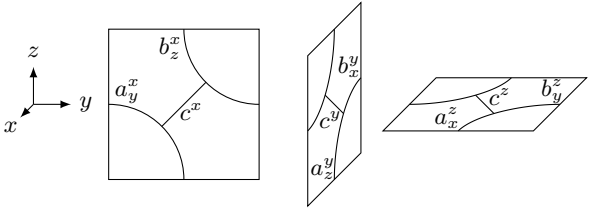

\begin{figure}[t]
    \centering
    \includegraphics[width=\columnwidth]{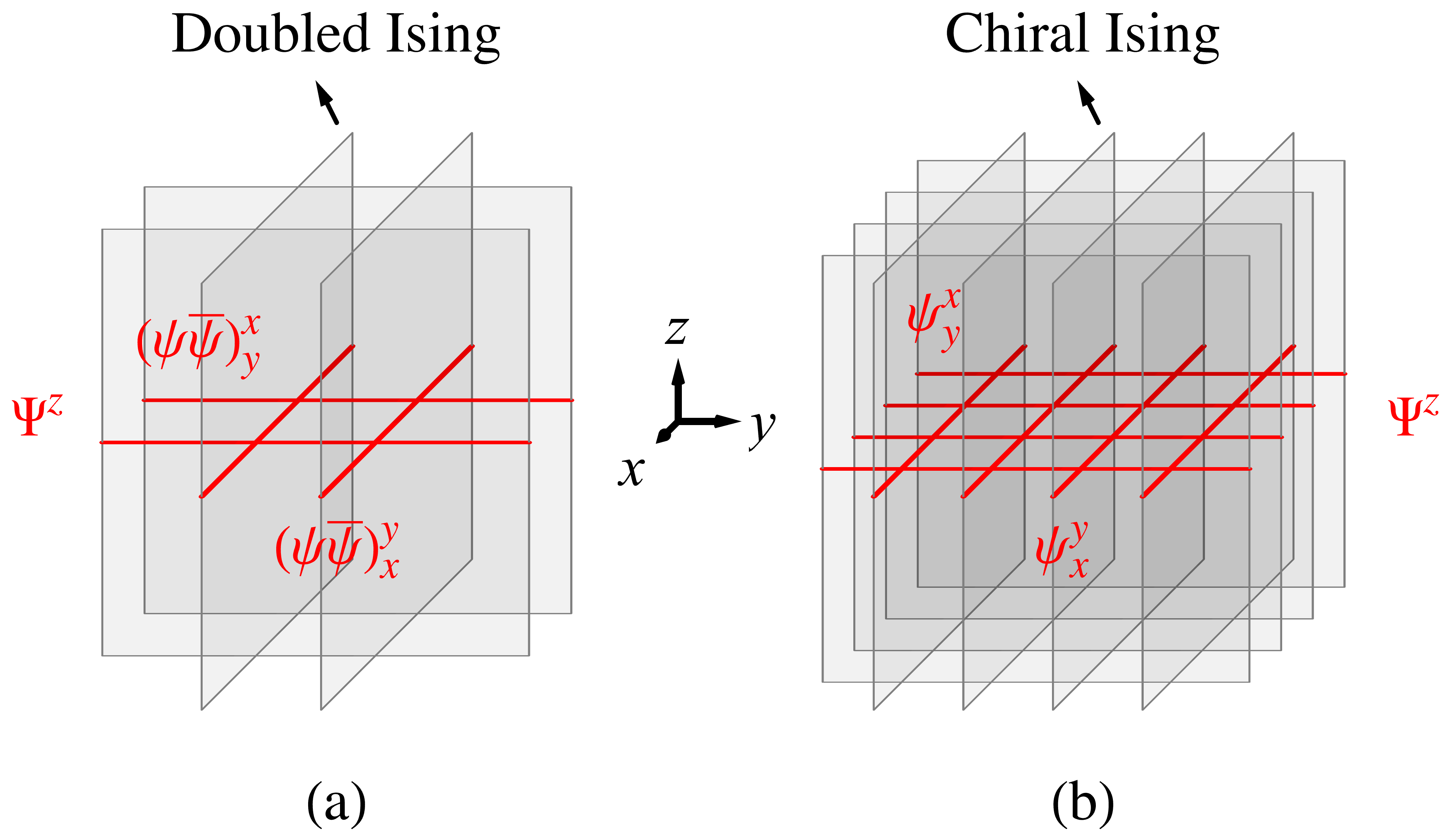}
    \captionsetup{justification=Justified}
    \caption{Net-shaped logical operator $\Psi^z$ defined in \eqref{eq:net_def}, which is to be condensed in Ising cage-net. In (a), each plane is a layer of doubled Ising, and the red strings are $(\psi\bar\psi)^x_y(i)$ and $(\psi\bar\psi)^y_x(j)$. In (b), equivalently, each plane is a layer of chiral Ising, and the red strings are $\psi^x_y(i)$ and $\psi^y_x(j)$.}
    \label{fig:net}
\end{figure}

\begin{figure}[t]
    \centering
    \includegraphics[scale = 0.1]{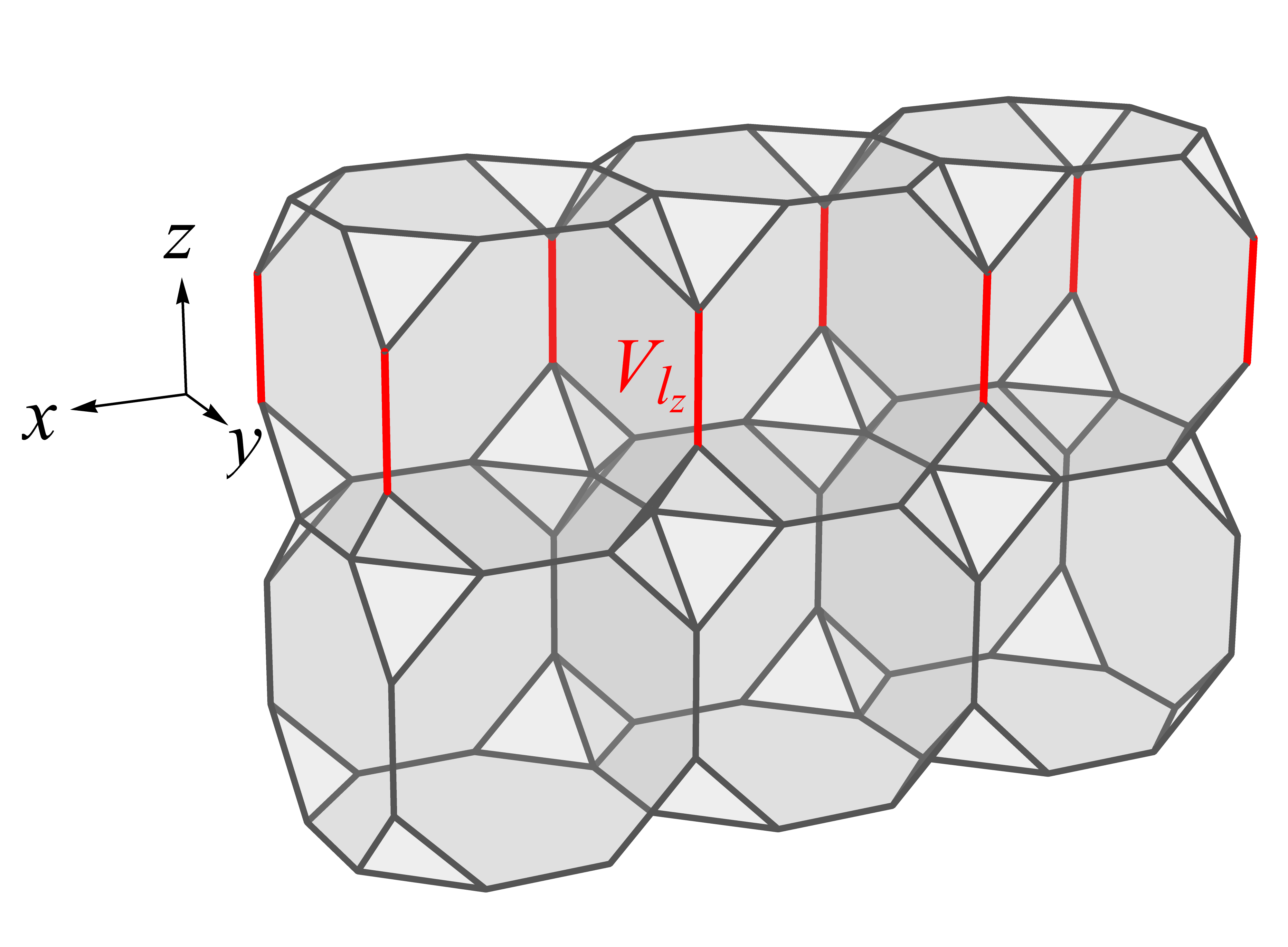}
    \captionsetup{justification=Justified}
    \caption{Action on the lattice degrees of freedom of the operator $\Psi^z$, which is to be condensed in Ising cage-net. The product of $V_{l_z}$ on the red edges (the set $T^z$ in \eqref{eq:net_def}) is the net-shaped logical operator $\Psi^z$ shown in Fig.~\ref{fig:net}~(a). Note that \eqref{eq:net_def} shown here is a logical operator, whereas \eqref{eq:1f_creation} shown in Fig.~\ref{fig:condense_1f} creates excitations.}
    \label{fig:condense_3f}
\end{figure}

We consider a system where we stack $L_x$, $L_y$ and $L_z$ layers of doubled Ising in the $x$, $y$ and $z$ directions, respectively. The elementary operators here are products of the $(2+1)$D elementary operators $v^x(a^x_y,b^x_z,c^x)$ in the $yz$ planes, $v^y(a^y_z,b^y_x,c^y)$ in the $zx$ planes, and $v^z(a^z_x,b^z_y,c^z)$ in the $xy$ planes (Fig.~\ref{fig:op3d}). We will also use notations such as $\psi^x_y(i)$ to denote the string operator of $\psi$ from the $i$th plane orthogonal to the $x$ direction (i.e.\ a $yz$ plane) traversing the $y$ direction. To obtain Ising cage-net from these decoupled layers, we need to condense $\psi\bar\psi$ p-loops as discussed in Section~\ref{sec:H}. Since our approach uses the operator algebra on the ground space, we need to combine the condensation operators $V_{l_\mu}$ defined in \eqref{eq:condense_op} into a logical operator (of the decoupled layers). An example of such a logical operator is shown in Fig.~\ref{fig:net}~(a), which looks like a ``net'' orthogonal to the $z$ direction. We call it a $\Psi$-net and denote it by $\Psi^z$. Explicitly, if $T^z$ is a set of principal edges $l_z$ related to each other by translation in the $x$ and $y$ directions (red edges in Fig.~\ref{fig:condense_3f}), then
\begin{equation}\label{eq:net_def}
    \Psi^z=\prod_{l_z\in T^z}V_{l_z}=\prod_{i=1}^{L_x}(\psi\bar\psi)^x_y(i)\prod_{j=1}^{L_y}(\psi\bar\psi)^y_x(j).
\end{equation}
Different choices of $T^z$ at different $xy$ planes give the same $\Psi^z$ when acting on the ground space. Similarly, we can define $\Psi^x$ and $\Psi^y$.

If we take the net shape of Fig.~\ref{fig:net}~(a) but replace all $\psi\bar\psi$'s with $\sigma\bar\sigma$'s, then we obtain an operator which we call a $\Sigma$-net, or $\Sigma^z$ in this case, to be more precise. Each $\Sigma^\alpha$ splits into two operators $\Sigma^\alpha=e^\alpha+m^\alpha$ of the same net shape. In the case of $\Sigma^z$, the operators $e^z$ and $m^z$ are distinguished by the parity $p^z$ of the fermion mode \[\prod_{i=1}^{L_x}(\sigma\bar\sigma)^x(i)\prod_{j=1}^{L_y}(\sigma\bar\sigma)^y(j),\] which is a good quantum number. This is because anyons such as $\sigma^x(i)$ which can change $p^z$ by braiding with $\Sigma^z$ are confined.

The semisimple algebra of the decoupled layers is \[A=(\text{Mat}_3\oplus\text{Mat}_1)^{\otimes2(L_x+L_y+L_z)}.\] Besides the condensation condition, we need to quotient $A$ by relations due to deconfined excitations. Since Ising cage-net has deconfined fractons, lineons and planons, it is not obvious where exactly the relations come from. Therefore, we return to the Hamiltonian \eqref{eq:H} and construct the relations from the Hamiltonian terms.

Firstly, the Hamiltonian \eqref{eq:H} contains the doubled Ising plaquette terms $B_p^0=1$ and $B_p^2$, so a ground state must satisfy the projector
\begin{equation}\label{eq:small_loop}
    \frac{1}{2}\left(1+B_p^2\right)=\frac{1}{2}\left(B_p^1\right)^2
\end{equation}
In the string-net model of doubled Ising, a 1-loop on a (smallest) plaquette can be viewed as a $\sigma$-loop or, equivalently, a $\bar\sigma$-loop. Here, we interpret \eqref{eq:small_loop} as creating a loop of $\sigma\bar\sigma$ at a plaquette. Suppose that this plaquette term is placed ``around the corner edges'' like
\[\frac{1}{2}\left(B_p^1\right)^2=
\vcenter{\hbox{
\begin{tikzpicture}
    \draw (-1, -1) rectangle (1, 1);
    \draw (-0.75, -1) arc (0:90:0.25);
    \draw (1, -0.75) arc (90:180:0.25);
    \draw (0.75, 1) arc (180:270:0.25);
    \draw (-1, 0.75) arc (270:360:0.25);
\end{tikzpicture}}}~.
\]
This simplifies to the relation
\begin{equation}\label{eq:rel_plaquette}
    r^\alpha(i)\bar r^\alpha(i)=1
\end{equation}
in each layer~$i$ orthogonal to the $\alpha$ direction, where e.g. \[r^x(i)=\frac{1}{2}\left(1+\psi^x_y(i)+\psi^x_z(i)-\psi^x_y(i)\psi^x_z(i)\right),\] and similarly for $\bar r^\alpha(i)$.

\begin{figure}[t]
    \centering
    \captionsetup{justification=Justified}
    \begin{tikzpicture}[baseline={([yshift=-.5ex]current bounding box.center)}, every node/.style={scale=1}]
        \begin{scope}
            \pgfmathsetmacro{\lineW}{0.4}
            \pgfmathsetmacro{\ClineW}{1}
            \pgfmathsetmacro{\framelineW}{0.4}
            \pgfmathsetmacro{\radius}{0.7}
            \pgfmathsetmacro{\halfSquareL}{\radius*cos(22.5)}
            \pgfmathsetmacro{\frameLen}{3}
            \pgfmathsetmacro{\xshift}{4.5}
            
            \tikzset{yzplane/.style={canvas is yz plane at x=#1,very thin}}
            \tikzset{xzplane/.style={canvas is xz plane at y=#1,very thin}}
            \tikzset{xyplane/.style={canvas is xy plane at z=#1,very thin}}
            
            \begin{scope}
                \draw[] (\xshift,0,0) node[] (FrameCorner) {}; 
                \draw[] (0-\frameLen+\xshift,0-\frameLen,0-\frameLen) node[] (FrameCornerDashed) {}; 
                
                \begin{scope}
                    \draw[line width = \framelineW, dashed] (FrameCornerDashed.center) -- ++ (0,\frameLen,0);
                    \draw[line width = \framelineW, dashed] (FrameCornerDashed.center) -- ++ (\frameLen,0,0);
                    \draw[line width = \framelineW, dashed] (FrameCornerDashed.center) -- ++ (0,0,\frameLen);
                \end{scope}
                
                \begin{scope}
                    \begin{scope}
                        \begin{scope}[xyplane=-\halfSquareL]
                            \clip (-\frameLen+\xshift,-\frameLen) rectangle (0+\xshift,0);
                            \draw[draw = green, line width = \ClineW] (-\frameLen+\xshift,-0.5*\frameLen) circle (0.7*\radius); 
                        \end{scope}
                        \begin{scope}[xzplane=-0.5*\frameLen-\halfSquareL]
                            \clip (-\frameLen+\xshift,0) rectangle (0+\xshift,-\frameLen);
                            \draw[draw = red, line width = \ClineW] (-\frameLen+\xshift,0)  circle  (0.7*\radius); 
                        \end{scope}
                    \end{scope}
                    
                    \begin{scope}
                        \begin{scope}[xyplane=-\frameLen+\halfSquareL]
                            \clip (-\frameLen+\xshift,-\frameLen) rectangle (0+\xshift,0);
                            \draw[draw = green, line width = \ClineW] (-\frameLen+\xshift,-0.5*\frameLen) circle (0.7*\radius); 
                        \end{scope}
                        \begin{scope}[yzplane=-\frameLen+\xshift+\halfSquareL]
                            \clip (-\frameLen,0) rectangle (0,-\frameLen);
                            \draw[draw = blue, line width = \ClineW] (-0.5*\frameLen,0) circle  (0.7*\radius); 
                        \end{scope}
                        \begin{scope}[xzplane=-0.5*\frameLen+\halfSquareL]
                            \clip (-\frameLen+\xshift,0) rectangle (0+\xshift,-\frameLen);
                            \draw[draw = red, line width = \ClineW] (-\frameLen+\xshift,0) circle  (0.7*\radius); 
                        \end{scope}
                    \end{scope}
                \end{scope}
                
                \begin{scope}
                    \begin{scope}
                        \begin{scope}[xzplane=-0.5*\frameLen-\halfSquareL]
                            \clip (-\frameLen+\xshift,0) rectangle (0+\xshift,-\frameLen);
                            \draw[draw = red, line width = \ClineW] (-\frameLen+\xshift,-\frameLen)  circle  (0.7*\radius); 
                        \end{scope}
                    \end{scope}
                    
                    \begin{scope}
                        \begin{scope}[yzplane=-\frameLen+\xshift+\halfSquareL]
                            \clip (-\frameLen,0) rectangle (0,-\frameLen);
                            \draw[draw = blue, line width = \ClineW] (-0.5*\frameLen,-\frameLen) circle  (0.7*\radius); 
                        \end{scope}
                        \begin{scope}[xzplane=-0.5*\frameLen+\halfSquareL]
                            \clip (-\frameLen+\xshift,0) rectangle (0+\xshift,-\frameLen);
                            \draw[draw = red, line width = \ClineW] (-\frameLen+\xshift,-\frameLen) circle  (0.7*\radius); 
                        \end{scope}
                    \end{scope}
                \end{scope}
                
                \begin{scope}
                    \begin{scope}
                        \begin{scope}[yzplane=\xshift-\halfSquareL]
                            \clip (-\frameLen,0) rectangle (0,-\frameLen);
                            \draw[draw = blue, line width = \ClineW] (-0.5*\frameLen,-\frameLen) circle  (0.7*\radius); 
                        \end{scope}
                        \begin{scope}[xzplane=-0.5*\frameLen-\halfSquareL]
                            \clip (-\frameLen+\xshift,0) rectangle (0+\xshift,-\frameLen);
                            \draw[draw = red, line width = \ClineW] (\xshift,-\frameLen)  circle  (0.7*\radius); 
                        \end{scope}
                    \end{scope}
                    
                    \begin{scope}
                        \begin{scope}[xyplane=-\frameLen+\halfSquareL]
                            \clip (-\frameLen+\xshift,-\frameLen) rectangle (0+\xshift,0);
                            \draw[draw = green, line width = \ClineW] (\xshift,-0.5*\frameLen) circle (0.7*\radius); 
                        \end{scope}
                        \begin{scope}[xzplane=-0.5*\frameLen+\halfSquareL]
                            \clip (-\frameLen+\xshift,0) rectangle (0+\xshift,-\frameLen);
                            \draw[draw = red, line width = \ClineW] (\xshift,-\frameLen) circle  (0.7*\radius); 
                        \end{scope}
                    \end{scope}
                \end{scope}
                
                \begin{scope}
                    \begin{scope}
                        \begin{scope}[xyplane=-\halfSquareL]
                            \clip (-\frameLen+\xshift,-\frameLen) rectangle (\xshift,0);
                            \draw[draw = green, line width = \ClineW] (\xshift,-0.5*\frameLen) circle (0.7*\radius); 
                        \end{scope}
                        \begin{scope}[yzplane=\xshift-\halfSquareL]
                            \clip (-\frameLen,0) rectangle (0,-\frameLen);
                            \draw[draw = blue, line width = \ClineW] (-0.5*\frameLen,0) circle  (0.7*\radius); 
                        \end{scope}
                        \begin{scope}[xzplane=-0.5*\frameLen-\halfSquareL]
                            \clip (-\frameLen+\xshift,0) rectangle (0+\xshift,-\frameLen);
                            \draw[draw = red, line width = \ClineW] (\xshift,0)  circle  (0.7*\radius); 
                        \end{scope}
                    \end{scope}
                    
                    \begin{scope}
                        \begin{scope}[xzplane=-0.5*\frameLen+\halfSquareL]
                            \clip (-\frameLen+\xshift,0) rectangle (0+\xshift,-\frameLen);
                            \draw[draw = red, line width = \ClineW] (\xshift,0) circle  (0.7*\radius); 
                        \end{scope}
                    \end{scope}
                \end{scope}
                
                \begin{scope}
                    \draw[draw = black, line width = \framelineW] (FrameCorner.center) -- ++(0,0,-\frameLen) -- ++ (0,-\frameLen,0) -- ++ (0,0,\frameLen) -- cycle; 
                    \draw[draw = black, line width = \framelineW] (FrameCorner.center) -- ++(-\frameLen,0,0) -- ++ (0,-\frameLen,0) -- ++ (\frameLen,0,0) -- cycle; 
                    \draw[draw = black, line width = \framelineW] (FrameCorner.center) -- ++(-\frameLen,0,0) -- ++ (0,0,-\frameLen) -- ++ (\frameLen,0,0) -- cycle; 
                \end{scope}
            \end{scope}
            
            \pgfmathsetmacro{\CordLen}{0.4}
            \draw[] (0.1*\frameLen,-0.5*\frameLen,0) node[] (CordCenter) {}; 
            \draw[-latex] (CordCenter.center) -- ++ (1.5*\CordLen,0,0); 
            \draw[-latex] (CordCenter.center) -- ++ (0,0,1.5*\CordLen); 
            \draw[-latex] (CordCenter.center) -- ++ (0,1.5*\CordLen,0); 
            \node [] at (0.1*\frameLen+2*\CordLen,-0.5*\frameLen,0) {$y$};
            \node [] at (0.1*\frameLen,-0.5*\frameLen,2.2*\CordLen) {$x$};
            \node [] at (0.1*\frameLen,-0.5*\frameLen+2*\CordLen,0) {$z$};
        \end{scope}
    \end{tikzpicture}
    \caption{Cage term $B_c$ of Ising cage-net placed 'around the corner edges'. The red, green and blue strings are 1-loops in the $xy$, $yz$ and $zx$ planes, respectively.}
    \label{fig:cage_relation}
\end{figure}
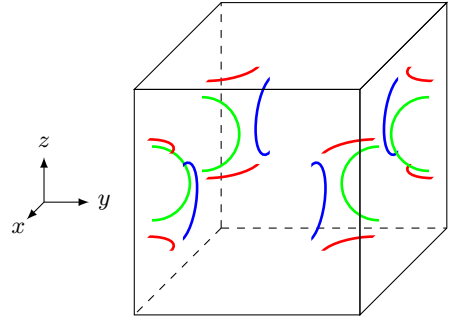

Secondly, we can also place a cage term $B_c$ ``around the corner edges'' (Fig.~\ref{fig:cage_relation}). This term involves 1-loops in the $xy$, $yz$ and $zx$ planes. In the setup of Fig.~\ref{fig:cage_relation}, we can bring the 1-loops closer together by enlarging the cube $c$ to size $L_x\times L_y\times 1$. The result is a flat, degenerate cuboid, some of whose edges coincide with each other. This enlargement is allowed since the 1-loops can be deformed individually in each layer of doubled Ising and the enlarged cage term commutes with the condensation terms $V_{l_\mu}$. We can then simplify this large cage term. The red strings give
\[
\vcenter{\hbox{
    \begin{tikzpicture}
        \draw (-1, -1) rectangle (1, 1);
        \draw[red] (-1, 0.1) -- (-0.35, 0.1);
        \draw[red] (-1, -0.1) -- (-0.35, -0.1);
        \draw[red] (1, 0.1) -- (0.35, 0.1);
        \draw[red] (1, -0.1) -- (0.35, -0.1);
        \draw[red] (-0.1, -1) -- (-0.1, -0.35);
        \draw[red] (0.1, -1) -- (0.1, -0.35);
        \draw[red] (-0.1, 1) -- (-0.1, 0.35);
        \draw[red] (0.1, 1) -- (0.1, 0.35);
        \draw[red] (-0.1, -0.35) arc (0:90:0.25);
        \draw[red] (0.35, -0.1) arc (90:180:0.25);
        \draw[red] (0.1, 0.35) arc (180:270:0.25);
        \draw[red] (-0.35, 0.1) arc (270:360:0.25);
        
        \draw[red] (-1, 0.2) -- (-0.35, 0.2);
        \draw[red] (-1, -0.2) -- (-0.35, -0.2);
        \draw[red] (1, 0.2) -- (0.35, 0.2);
        \draw[red] (1, -0.2) -- (0.35, -0.2);
        \draw[red] (-0.2, -1) -- (-0.2, -0.35);
        \draw[red] (0.2, -1) -- (0.2, -0.35);
        \draw[red] (-0.2, 1) -- (-0.2, 0.35);
        \draw[red] (0.2, 1) -- (0.2, 0.35);
        \draw[red] (-0.2, -0.35) arc (0:90:0.15);
        \draw[red] (0.35, -0.2) arc (90:180:0.15);
        \draw[red] (0.2, 0.35) arc (180:270:0.15);
        \draw[red] (-0.35, 0.2) arc (270:360:0.15);
        
        \draw[->, -latex] (-2, 0) -- (-1.5, 0);
        \draw[->, -latex] (-2, 0) -- (-2, 0.5);
        \node at (-1.3, 0) {$x$};
        \node at (-2, 0.7) {$y$};
    \end{tikzpicture}}}\ =2r^z(i)r^z(i+1),
\]
where the two 1-loops are in different $xy$ planes but drawn in the same plane for illustration, and we draw the degenerate cuboid as a large yet non-degenerate one. We chose to interpret the two 1-loops as two $\sigma$-loops; other interpretations such as one $\sigma$-loop and one $\bar\sigma$-loop are all equivalent due to \eqref{eq:rel_plaquette}. The green strings give
\[
\vcenter{\hbox{
    \begin{tikzpicture}
        \draw (-1, -1) rectangle (1, 1);
        \draw[green] (-1, 0.25) -- (-0.35, 0.25);
        \draw[green] (-1, -0.25) -- (-0.35, -0.25);
        \draw[green] (1, 0.25) -- (0.35, 0.25);
        \draw[green] (1, -0.25) -- (0.35, -0.25);
        \draw[green] (-0.35, -0.25) arc (-90:90:0.25);
        \draw[green] (0.35, 0.25) arc (90:270:0.25);
        
        \draw[green] (-1, 0.15) -- (-0.35, 0.15);
        \draw[green] (-1, -0.15) -- (-0.35, -0.15);
        \draw[green] (1, 0.15) -- (0.35, 0.15);
        \draw[green] (1, -0.15) -- (0.35, -0.15);
        \draw[green] (-0.35, -0.15) arc (-90:90:0.15);
        \draw[green] (0.35, 0.15) arc (90:270:0.15);
        
        \draw[->, -latex] (-2, 0) -- (-1.5, 0);
        \draw[->, -latex] (-2, 0) -- (-2, 0.5);
        \node at (-1.3, 0) {$y$};
        \node at (-2, 0.7) {$z$};
    \end{tikzpicture}}}\ =2,
\]
Note that this simplification uses only the fusion rules, $F$-symbols and $R$-symbols. Similarly, the blue strings simplify to a constant 2. Therefore, Fig.~\ref{fig:cage_relation} gives a relation $r^z(i)r^z(i+1)=1$.

In summary, the Hamiltonian \eqref{eq:H} implies that the product of $r^\alpha(i)$ or $\bar r^\alpha(i)$ with any other $r^\alpha(j)$ or $\bar r^\alpha(j)$ should be 1, where $i$ and $j$ may or may not be equal. We observe that for the purpose of writing down relations, there is no difference between anyons with and without bars. Therefore, from now on we will consider the system as $2L_x$, $2L_y$ and $2L_z$ layers of chiral Ising. The names of operators change accordingly, e.g. \[\Psi^z=\prod_{i=1}^{2L_x}\psi^x_y(i)\prod_{j=1}^{2L_y}\psi^y_x(j),\] as shown in Figure~\ref{fig:net}~(b). Let $M$ be the subalgebra of $A$ generated by $\Psi^x$, $\Psi^y$ and $\Psi^z$, and $M'$ the commutant of $M$. Inside $M'$, the relations amount to central projectors
\begin{equation}\label{eq:condense_ICN}
    P_\text{c}=\frac{1}{8}(1+\Psi^x)(1+\Psi^y)(1+\Psi^z)
\end{equation}
due to condensation, and
\begin{equation}\label{eq:pair_ICN}
    P^\alpha_{ij}=\frac{1}{2}\left(1+r^\alpha(i)r^\alpha(j)\right)
\end{equation}
due to deconfined planons and cage terms. Their product is
\begin{equation}\label{eq:proj_ICN}
    P=P_c\prod_\alpha\prod_{i<j}P^\alpha_{ij}.
\end{equation}
With the above setup, we are ready to calculate the GSD.

\subsection{GSD from Cartan subalgebra}
\label{sec:ICN_Cartan}

Following Section~\ref{sec:1f_Cartan}, we calculate the GSD of Ising cage-net using a Cartan subalgebra. The semisimple algebra $A$ has a Cartan subalgebra $C$ spanned by the elementary operators with no $\sigma$. Just like in Section~\ref{sec:1f_Cartan}, it happens that $C\subset M'$, and the central projectors $P_\text{c}$ and $P^\alpha_{ij}$ all map $C$ to $C$. We also have the splitting of the $\Sigma$-nets, but this does not enlarge the Cartan subalgebra. This is because every $\Sigma^\alpha$ (and hence $e^\alpha$ and $m^\alpha$) braids non-trivially with some $\psi$ operator. Therefore, we have $\text{GSD}=\text{tr}(P)$, where the underlying vector space is $C$. Again using the argument in Section~\ref{sec:1f_Cartan}, if $P$ is expanded into a linear combination of elementary operators, then only the constant term $\mu_0=\mu(1,1,1,1,1,1)$ (which is called $\mu(1,1)$ for 1-F Ising) contributes to $\text{tr}(P)$.

To compute $\mu_0$, we need to expand \eqref{eq:proj_ICN}. This is very similar to the calculation in Section~\ref{sec:1f_Cartan}. Firstly, we have \[\prod_{i<j}P^\alpha_{ij}=\frac{1}{2^{2L_\alpha}}\left[\prod_{i=1}^{2L_\alpha}(1+r^\alpha(i))+\prod_{i=1}^{2L_\alpha}(1-r^\alpha(i))\right].\] Thus
\begin{align*}
    P=\frac{1}{8}(&1+\Psi^x+\Psi^y+\Psi^z\\
    &+\Psi^y\Psi^z+\Psi^z\Psi^x+\Psi^x\Psi^y+\Psi^x\Psi^y\Psi^z)\\
    &\times\prod_\alpha\frac{1}{2^{2L_\alpha}}\left[\prod_{i=1}^{2L_\alpha}(1+r^\alpha(i))+\prod_{i=1}^{2L_\alpha}(1-r^\alpha(i))\right].
\end{align*}
We need to find terms in the expansion of $\prod_\alpha(\cdots)$ that combines with one of the eight terms $1,\Psi^x,\ldots,\Psi^x\Psi^y\Psi^z$ to give a constant term. Now, for example, the only four terms in the expansion of $\prod_i(1+r^z(i))$ that can possibly contribute to $\mu_0$ are
\begin{align*}
    \left(\frac{3}{2}\right)^{2L_z},\ \prod_i\left(\frac{1}{2}\psi^z_x(i)\right),&\ \prod_i\left(\frac{1}{2}\psi^z_y(i)\right),\\
    \text{and }&\prod_i\left(-\frac{1}{2}\psi^z_x(i)\psi^z_y(i)\right).
\end{align*}
Therefore, we can write
\begin{widetext}
\begin{align*}
    P=\frac{1}{8}(&1+\Psi^x+\Psi^y+\Psi^z+\Psi^y\Psi^z+\Psi^z\Psi^x+\Psi^x\Psi^y+\Psi^x\Psi^y\Psi^z)\\
    &\times\frac{1}{2^{4L_x}}\left[\left(9^{L_x}+1\right)+2\prod_i\psi^x_y(i)+2\prod_i\psi^x_z(i)+2\prod_i\psi^x_y(i)\psi^x_z(i)\right]\\
    &\times\frac{1}{2^{4L_y}}\left[\left(9^{L_y}+1\right)+2\prod_j\psi^y_z(j)+2\prod_j\psi^y_x(j)+2\prod_j\psi^y_z(j)\psi^y_x(j)\right]\\
    &\times\frac{1}{2^{4L_z}}\left[\left(9^{L_z}+1\right)+2\prod_k\psi^z_x(k)+2\prod_k\psi^z_y(k)+2\prod_k\psi^z_x(k)\psi^z_y(k)\right]+\cdots,
\end{align*}
\end{widetext}
where ``$\cdots$'' means terms that cannot possibly contribute to $\mu_0$. Up to permutation of $x$, $y$ and $z$, the pairing of the terms works as follows:
\begingroup
\allowdisplaybreaks
\begin{align*}
    1\iff&\left(9^{L_x}+1\right)\times\left(9^{L_y}+1\right)\times\left(9^{L_z}+1\right),\\
    \Psi^z\iff&\left(9^{L_z}+1\right)\times2\prod_i\psi^x_y(i)\times2\prod_j\psi^y_x(j),\\
    \Psi^x\Psi^y\iff&2\prod_i\psi^x_z(i)\times 2\prod_j\psi^y_z(j)\\
    &\times 2\prod_k\psi^z_x(k)\psi^z_y(k),\\
    \Psi^x\Psi^y\Psi^z\iff&2\prod_i\psi^x_y(i)\psi^x_z(i)\times 2\prod_j\psi^y_z(j)\psi^y_x(j)\\
    &\times 2\prod_k\psi^z_x(k)\psi^z_y(k),
\end{align*}
\endgroup
where ``$\iff$'' indicates the pairing. Combining these together, we obtain
\begin{align*}
    \text{GSD}=\ & 2^{4(L_x+L_y+L_z)}\mu_0\\
    =\ &\frac{1}{8}\bigl[\left(9^{L_x}+1\right)\left(9^{L_y}+1\right)\left(9^{L_z}+1\right)\\
    &\quad+4\left(9^{L_x}+1\right)+4\left(9^{L_y}+1\right)+4\left(9^{L_z}+1\right)\\
    &\quad+8+8+8+8\bigr]\\
    =\ &\frac{1}{8}(E_3+E_2+5E_1+45),
\end{align*}
where $E_3 = 9^{L_x+L_y+L_z}$, $E_2 = 9^{L_x+L_y}+9^{L_y+L_z}+9^{L_z+L_x}$, and $E_1 = 9^{L_x}+9^{L_y}+9^{L_z}$. In Appendix~\ref{app:small_size}, we confirm this result for the smallest system size $L_x=L_y=L_z=1$ with a lattice calculation independent of the operator algebra approach.

\subsection{GSD from full algebra}
\label{sec:ICN_full}

To conclude our discussion of Ising cage-net, we calculate its GSD using the full semisimple algebra, as we did in Sections~\ref{sec:SN} and \ref{sec:1f_full}.

Similar to 1-F Ising, the central projectors $P^\alpha_{ij}$ defined in \eqref{eq:pair_ICN} kill most of the components of the semisimple algebra $M'$. This is because for each $\alpha$, projection by $P^\alpha_{ij}$ forces the $r^\alpha(i)$'s to be all $+1$ or all $-1$. We have \[\prod_\alpha\prod_{i<j}P^\alpha_{ij}A=(B^x_0\oplus B^x_1)\otimes(B^y_0\oplus B^y_1)\otimes(B^z_0\oplus B^z_1),\] where $B^\alpha_0\cong\text{Mat}_{9^{L_\alpha}}$ and $B^\alpha_1\cong\text{Mat}_1$. We can define central projectors \[Q_{s_xs_ys_z}=\prod_\alpha\left[\frac{1}{2^{2L_\alpha}}\prod_{i=1}^{2L_\alpha}\left(1+(-1)^{s_\alpha}r^\alpha(i)\right)\right],\] where $s_\alpha=0$ or 1, which project onto the components \[B_{s_xs_ys_z}=\bigotimes_\alpha B^\alpha_{s_\alpha}.\] We need to find the action of $P_\text{c}$ defined in \eqref{eq:condense_ICN} on \[\left[\bigotimes_\alpha(B^\alpha_0\oplus B^\alpha_1)\right]\cap M'.\] This intersection has eight components, which are $B_{000}\cap M'$ and so on. Up to permutation of $x$, $y$ and $z$, we have four cases, and we discuss them in ascending order of difficulty:

(1) On $B_{111}\cap M'=B_{111}$, every $\psi^\alpha_\beta(i)$ acts as $-1$, so $P_\text{c}Q_{111}M'=B_{111}=\text{Mat}_1$.

(2) On $B_{110}\cap M'$, each of $\psi^x_y(i)$, $\psi^x_z(i)$, $\psi^y_z(j)$ and $\psi^y_x(j)$ acts as $-1$, while each of $\psi^z_x(k)$ and $\psi^z_y(k)$ has the representation $\rho_3$ given by \eqref{eq:matrix3}. By Lemma~\ref{thm:center}, the central projector $P_\text{c}Q_{110}$ is primitive. To determine the matrix algebra $P_\text{c}Q_{110}M'$, we need the representation $\rho_l$ of $\text{Mat}_1\otimes\text{Mat}_1\otimes\text{Mat}_{9^{L_z}}$ where $l=9^{L_z}$. More precisely, we need the common eigenspace $\{w\}$ such that $\rho_l(\Psi^\alpha)w=+w$ for all $\alpha$. Now we already have $\rho_l(\Psi^z)w=+w$ because $\psi^x_y(i)=-1$ and $\psi^y_x(j)=-1$. To ensure e.g.\ $\rho_l(\Psi^x)w=+w$, we must have
\begin{equation}\label{eq:eigs_1}
    \left[\bigotimes_k\rho_3[\psi^z_y(k)]\right]w=+w,
\end{equation}
since $\psi^y_z(j)=-1$. Similarly, we must also have
\begin{equation}\label{eq:eigs_2}
    \left[\bigotimes_k\rho_3[\psi^z_x(k)]\right]w=+w.
\end{equation}
The dimension of the eigenspace that satisfies \eqref{eq:eigs_1} and \eqref{eq:eigs_2} is precisely $D^{++}_{2L_z}$ defined in \eqref{eq:common_eigen}. Therefore, we have $P_\text{c}Q_{110}M'=\text{Mat}_{(9^{L_z}+3)/4}$.

(3) On $B_{001}\cap M'$, each of $\psi^z_x(k)$ and $\psi^z_y(k)$ acts as $-1$, while each of $\psi^x_y(i)$, $\psi^x_z(i)$, $\psi^y_z(j)$ and $\psi^y_x(j)$ has the representation $\rho_3$. To determine the matrix algebra $P_\text{c}Q_{001}M'$, we need the common eigenspace $\{w\}$ such that $\rho_l(\Psi^\alpha)w=+w$ in the representation $\rho_l$ of $\text{Mat}_{9^{L_x}}\otimes\text{Mat}_{9^{L_y}}\otimes\text{Mat}_1$ where $l=9^{L_x+L_y}$. From $\rho_l(\Psi^x)w=\rho_l(\Psi^y)w=+w$ and $\psi^z_x(k)=\psi^z_y(k)=-1$ we obtain \[\left[\bigotimes_i\rho_3[\psi^x_z(i)]\right]w=\left[\bigotimes_j\rho_3[\psi^y_z(j)]\right]w=+w.\] Meanwhile, $\rho_l(\Psi^z)w=+w$ implies two possibilities
\begin{equation}\label{eq:eigs_3}
    \left[\bigotimes_i\rho_3[\psi^x_y(i)]\right]w=\left[\bigotimes_j\rho_3[\psi^y_x(i)]\right]w=\pm w.
\end{equation}
If we take the $+w$ in \eqref{eq:eigs_3}, then we obtain a subspace of dimension $D^{++}_{2L_x}D^{++}_{2L_y}$. On the other hand, if we take the $-w$ in \eqref{eq:eigs_3}, then we obtain a subspace of dimension $D^{-+}_{2L_x}D^{+-}_{2L_y}$. Overall, we have $P_\text{c}Q_{001}M'=\text{Mat}_{(9^{L_x+L_y}+9^{L_x}+9^{L_y}+5)/8}$.

(4) On $B_{000}\cap M'$, every $\psi^\alpha_\beta(i)$ has the representation $\rho_3$. To determine the matrix algebra $P_\text{c}Q_{000}M'$, we need the common eigenspace $\{w\}$ such that $\rho_l(\Psi^\alpha)w=+w$ in the representation of $\text{Mat}_{9^{L_x}}\otimes\text{Mat}_{9^{L_y}}\otimes\text{Mat}_{9^{L_z}}$ where $l=9^{L_x+L_y+L_z}$. This gives the equations
\begin{align}
    \left[\bigotimes_j\rho_3[\psi^y_z(j)]\right]w=\left[\bigotimes_k\rho_3[\psi^z_y(k)]\right]w&=\pm w, \label{eq:eigs4}\\
    \left[\bigotimes_k\rho_3[\psi^z_x(k)]\right]w=\left[\bigotimes_i\rho_3[\psi^x_z(i)]\right]w&=\pm w, \label{eq:eigs5}\\
    \left[\bigotimes_i\rho_3[\psi^x_y(i)]\right]w=\left[\bigotimes_j\rho_3[\psi^y_x(i)]\right]w&=\pm w. \label{eq:eigs6}
\end{align}
Depending on the choice of $\pm w$ in these equations, we have eight possibilities. For example, we can choose $-w$ in \eqref{eq:eigs4} and \eqref{eq:eigs5} and $+w$ in \eqref{eq:eigs6}, which has a contribution of $D^{+-}_{2L_x}D^{-+}_{2L_y}D^{--}_{2L_z}$ to the dimension of the common eigenspace. The total dimension is
\begin{align*}
    &D^{++}_{2L_x}D^{++}_{2L_y}D^{++}_{2L_z}+\left(D^{-+}_{2L_x}D^{+-}_{2L_y}D^{++}_{2L_z}+\text{perm.}\right)\\
    +&\left(D^{+-}_{2L_x}D^{-+}_{2L_y}D^{--}_{2L_z}+\text{perm.}\right)+D^{--}_{2L_x}D^{--}_{2L_y}D^{--}_{2L_z}\\
    &\qquad\qquad\qquad=\frac{1}{8}(E_3+E_1+4),
\end{align*}
where ``perm.'' means permutations of $x$, $y$ and $z$. Since $D^{+-}_{2L}=D^{-+}_{2L}$, only cyclic permutations are included. Therefore, we have $P_\text{c}Q_{000}M'=\text{Mat}_{(E_3+E_1+4)/8}$.

Summarizing all four cases, we have
\begin{align*}
    PM'=\ &\text{Mat}_{(E_3+E_1+4)/8}\\
    &\oplus\left(\text{Mat}_{(9^{L_x+L_y}+9^{L_x}+9^{L_y}+5)/8}\oplus\text{perm.}\right)\\
    &\oplus\left(\text{Mat}_{(9^{L_z}+3)/4}\oplus\text{perm.}\right)\oplus\text{Mat}_1.
\end{align*}
Using Protocol~\ref{thm:conjecture} with the conjecture of ``filling the blanks'', we obtain $\text{GSD}=(E_3+E_2+5E_1+45)/8$. In Appendix~\ref{app:small_size}, we confirm this result for the smallest system size $L_x=L_y=L_z=1$ with a lattice calculation independent of the operator algebra approach.

\section{Summary}
\label{sec:summary}

In this paper, we have found the GSD of Ising cage-net to be \[\text{GSD}=\frac{1}{8}(E_3+E_2+5E_1+45),\] where $E_3 = 9^{L_x+L_y+L_z}$, $E_2 = 9^{L_x+L_y}+9^{L_y+L_z}+9^{L_z+L_x}$, and $E_1 = 9^{L_x}+9^{L_y}+9^{L_z}$. Based on this result, we have concluded that the Ising cage-net model cannot have a foliation structure as defined in Ref.~\onlinecite{Shirley2018}, because the GSD does not grow by integer multiples as the system size grows. On the other hand, we find that the foliation idea can be generalized to accommodate the renormalization group transformation of the Ising cage-net model. We discuss this generalized foliation in a separate paper\cite{https://doi.org/10.48550/arxiv.2301.00103}.

To compute the GSD, we have developed a collection of mathematical tools which we call the ``operator algebra approach''. In this approach, we view the ground space operator algebra $A_0$ of a topological or fractonic order as more fundamental than the ground space $\mathcal{H}_0$, and write $A_0$ as a semisimple algebra $A$ quotienting out some relations. In Protocols~\ref{thm:summary} and \ref{thm:conjecture}, we have outlined how this approach can be used to find $A_0$ and, with the operation of ``filling the blanks'', understand how boson condensation happens within this framework. The validity of this approach has been checked in some simple examples, namely chiral Ising in Section~\ref{sec:chiral}, doubled Ising condensed into the toric code in Section~\ref{sec:SN}, and 1-F Ising in Section~\ref{sec:1f-ICN}.

It may seem that the operator algebra approach is simply a trick for computing the GSD and, in particular, that the semisimple algebra $A$ is just an intermediate step in the calculation of the matrix algebra $A_0$. However, Protocol~\ref{thm:conjecture} suggests that $A$ has its own significance: When studying the condensation of $\psi\bar\psi$ in doubled Ising (Section~\ref{sec:SN}), the components $\text{Mat}_9$ and $\text{Mat}_1$ of $A$ are both important since they both intersect non-trivially with $M'$. If we focused only on the matrix algebra $\text{Mat}_9$ of doubled Ising, then we would miss the $\text{Mat}_1$. In another perspective, when constructing $A$, only the relations due to fusion rules, $F$-symbols and $R$-symbols are considered. Going from $A$ to $A_0$, we need to further quotient out the relations due to deconfined anyons. Now when bosons are condensed, certain anyons become confined. Such confinement reduces the number of relations due to deconfined anyons, but does not make any of the fusion rules, $F$-symbols or $R$-symbols invalid -- if they involve confined anyons then they do not affect $M$ or $M'$ anyway. In this sense, the relations have a ``hierarchy'', with relations due to deconfined anyons being ``less essential'' than relations due to fusion rules, $F$-symbols and $R$-symbols. Therefore, the operator algebra approach provides a new way of understanding topological and fractonic orders.

In fact, we can identify three important algebras in the operator algebra approach: $A_0$, $A$ and, in the context of condensation, $PM'$. Among them, the most physical one is $A_0$ since it is the actual algebra of logical operators. However, $A_0$ also contains the least information, since it can be derived from $A$ with the knowledge of operator relations, or from $PM'$ by ``filling the blanks''.

\begin{figure}[t]
    \centering
    \begin{tikzpicture}
        \draw (-4.2, -1.6) rectangle (-1, 1.6);
        
        \draw (-4.2, -0.2) -- (-1, -0.2);
        \draw (-4, 0) -- (-4, 0.2);
        \draw (-3.8, 0) -- (-3.8, 0.2);
        \draw (-3.6, 0) -- (-3.6, 0.2);
        \draw (-3.4, 0) -- (-3.4, 0.2);
        \draw (-3.2, 0) -- (-3.2, 0.2);
        \draw (-3, 0) -- (-3, 0.2);
        \draw (-2.8, 0) -- (-2.8, 0.2);
        \draw (-2.6, 0) -- (-2.6, 0.2);
        \draw (-2.4, 0) -- (-2.4, 0.2);
        \draw (-2.2, 0) -- (-2.2, 0.2);
        \draw (-2, 0) -- (-2, 0.2);
        \draw (-1.8, 0) -- (-1.8, 0.2);
        \draw (-1.6, 0) -- (-1.6, 0.2);
        \draw (-1.4, 0) -- (-1.4, 0.2);
        \draw (-1.2, 0) -- (-1.2, 0.2);
        
        \draw (-2.8, -1.6) -- (-2.8, 1.6);
        \draw (-2.6, -1.4) -- (-2.4, -1.4);
        \draw (-2.6, -1.2) -- (-2.4, -1.2);
        \draw (-2.6, -1) -- (-2.4, -1);
        \draw (-2.6, -0.8) -- (-2.4, -0.8);
        \draw (-2.6, -0.6) -- (-2.4, -0.6);
        \draw (-2.6, -0.4) -- (-2.4, -0.4);
        \draw (-2.6, -0.2) -- (-2.4, -0.2);
        \draw (-2.6, 0) -- (-2.4, 0);
        \draw (-2.6, 0.2) -- (-2.4, 0.2);
        \draw (-2.6, 0.4) -- (-2.4, 0.4);
        \draw (-2.6, 0.6) -- (-2.4, 0.6);
        \draw (-2.6, 0.8) -- (-2.4, 0.8);
        \draw (-2.6, 1) -- (-2.4, 1);
        \draw (-2.6, 1.2) -- (-2.4, 1.2);
        \draw (-2.6, 1.4) -- (-2.4, 1.4);
        
        \node[above] at (-1.4, 0.2) {$X_1$};
        \node[right] at (-2.4, 1.2) {$X_2$};
        \node[left] at (-2.8, -1.2) {$Z_1$};
        \node[below] at (-3.8, -0.2) {$Z_2$};
        
        \draw (0, -1.6) rectangle (3.2, 1.6);
        
        \draw (0, -0.2) -- (3.2, -0.2);
        \draw (0.2, 0) -- (0.2, 0.2);
        \draw (0.4, 0) -- (0.4, 0.2);
        \draw (0.6, 0) -- (0.6, 0.2);
        \draw (0.8, 0) -- (0.8, 0.2);
        \draw (1, 0) -- (1, 0.2);
        \draw (1.2, 0) -- (1.2, 0.2);
        \draw (1.4, 0) -- (1.4, 0.2);
        \draw (1.6, 0) -- (1.6, 0.2);
        \draw (1.8, 0) -- (1.8, 0.2);
        \draw (2, 0) -- (2, 0.2);
        \draw (2.2, 0) -- (2.2, 0.2);
        \draw (2.4, 0) -- (2.4, 0.2);
        \draw (2.6, 0) -- (2.6, 0.2);
        \draw (2.8, 0) -- (2.8, 0.2);
        \draw (3, 0) -- (3, 0.2);
        
        \draw (1.4, -1.6) -- (1.4, 1.6);
        \draw (1.6, -1.4) -- (1.8, -1.4);
        \draw (1.6, -1.2) -- (1.8, -1.2);
        \draw (1.6, -1) -- (1.8, -1);
        \draw (1.6, -0.8) -- (1.8, -0.8);
        \draw (1.6, -0.6) -- (1.8, -0.6);
        \draw (1.6, -0.4) -- (1.8, -0.4);
        \draw (1.6, -0.2) -- (1.8, -0.2);
        \draw (1.6, 0) -- (1.8, 0);
        \draw (1.6, 0.2) -- (1.8, 0.2);
        \draw (1.6, 0.4) -- (1.8, 0.4);
        \draw (1.6, 0.6) -- (1.8, 0.6);
        \draw (1.6, 0.8) -- (1.8, 0.8);
        \draw (1.6, 1) -- (1.8, 1);
        \draw (1.6, 1.2) -- (1.8, 1.2);
        \draw (1.6, 1.4) -- (1.8, 1.4);
        
        \node[above] at (2.8, 0.2) {$X_3$};
        \node[right] at (1.8, 1.2) {$X_4$};
        \node[left] at (1.4, -1.2) {$Z_3$};
        \node[below] at (0.4, -0.2) {$Z_4$};
    \end{tikzpicture}
    \captionsetup{justification=Justified}
    \caption{Two copies of the toric code and their logical operators. We use the setup of Ref.~\onlinecite{KITAEV20032} on a square lattice, which is not drawn explicitly.}
    \label{fig:tc}
\end{figure}
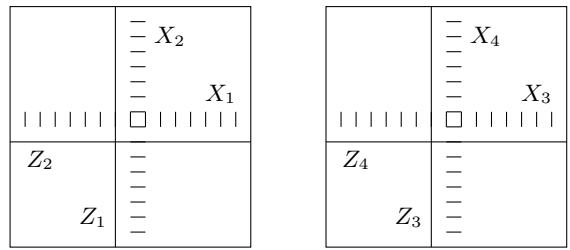

A direction for future work is to understand the operator algebra approach more systematically. At this moment, we can already see some advantages of this approach for studying fracton models: It does not care about spatial dimension; it handles logical operators of fully mobile particles, partially mobile particles and even non-point excitations (e.g.\ membrane operators) on equal ground; it also gives a very simple description of boson condensation. However, it is too simplistic to view the operator algebra approach as an \textit{abstract} semisimple algebra quotienting out some relations. For example, suppose that we have two copies of the toric code (Fig.~\ref{fig:tc}). The ground space of each copy is two qubits, say qubits~1 and 2 for the first copy, and qubits~3 and 4 for the second copy. Qubit~$i$ has logical operators $X_i$ and $Z_i$ which are Pauli matrices. If we condense $X_1$ and $X_2$, then we are left with the second copy of the toric code. Now suppose, instead, that we want to condense $X_1$ and $X_4$. On the one hand, this is unphysical, since enforcing $X_1=X_4=1$ leads to a non-robust ground space with infinite (i.e.\ extensive) degeneracy. This degeneracy can be lifted by local perturbations, and an example of such local perturbations is Pauli $X$'s on the horizontal edges of the first lattice and the vertical edges of the second lattice. The result is the trivial topological order. On the other hand, if $U$ is the (non-local) unitary that swaps qubits~2 and 4, then \[UX_1U^\dagger=X_1,\, UX_2U^\dagger=X_4.\] Thus from a purely abstract perspective, the pair of operators $(X_1,X_2)$ is the same as the pair of operators $(X_1,X_4)$. In other words, Protocol~\ref{thm:conjecture} tells us how to condense certain operators if we know that it is physical to do so, but it does not tell us which operators can be condensed physically. Therefore, we need to specify more information in the semisimple algebra $A$, especially regarding locality, in order to set up the mathematical structure in a physical way.

To help with this effort, we ask the following questions:
\begin{enumerate}[nolistsep, leftmargin=*]
    \item The matrix algebra $A_0$ is obtained from the semisimple algebra $A$ by quotienting out relations. Where do the relations come from in general? In Ising cage-net, we obtained the relations from the Hamiltonian, and they essentially say that a loop of a planon $\sigma^\alpha(i)\sigma^\alpha(j)$ should equal its quantum dimension. Alternatively, we can view the action of a cage term $B_c$ as lineon operators on the edges of a cube. However, if we want to interpret the relations from a lineon perspective, then it is unclear what analog of ``quantum dimension'' we should assign to the other side of the relation, because lineons cannot form contractible loops. For another example, in a $(3+1)$D gauge theory, a point charge can form contractible loops in the $xy$, $yz$ and $zx$ planes, giving three relations. Is this a general feature that depends on some notion of ``codimension'' of an elementary operator? It will be interesting to understand generally what kind of physical constraint we need to consider to derive all the necessary relations in $A$.
    
    \item Does the conjecture of ``filling the blanks'' in Protocol~\ref{thm:conjecture} hold for condensation transitions in general? Can the operation of ``filling the blanks'' be characterized more abstractly, e.g.\ by some universal property?
    
    \item How can the process of gauging, the opposite of condensation\cite{Davydov}, be understood in this approach?
    
    \item How about foliation and other notions of RG? What is an appropriate notion of equivalence here?
    
    \item Can we do reverse engineering, i.e.\ start with a matrix algebra written as a quotient of a semisimple algebra with some notion of locality, and construct a corresponding lattice model? Or even construct the spatial manifold without specifying it separately from the algebraic data?
\end{enumerate}
We hope that the operator algebra approach will shed light to the understanding of gapped orders of matter including topological and fractonic orders. Given that topological orders are characterized by modular tensor (higher) categories, if the operator algebra approach can characterize fractonic orders then it must be at least as sophisticated.

\begin{acknowledgments}
We are indebted to inspiring discussions with Bowen Yang, Kevin Slagle, Mike Hermele, Dave Aasen, and Meng Cheng. X.M, Z.-Y. W. and X.C. are supported by the National Science Foundation under award number DMR-1654340, the Simons collaboration on ``Ultra-Quantum Matter'' (grant number 651440), the Simons Investigator Award (award ID 828078) and the Institute for Quantum Information and Matter at Caltech. X.C. is also supported by the Walter Burke Institute for Theoretical Physics at Caltech. X.C. wants to thank the Institute for Advanced Study at Tsinghua University for hospitality when the paper was written.  Z.W. is partially supported by NSF grants
FRG-1664351, CCF 2006463, and ARO MURI contract W911NF-20-1-0082.  
\end{acknowledgments}

\bibliography{references}

\appendix

\section{Supplementary mathematics}
\label{app:math}

In this appendix, we discuss some mathematics supplementary to the main text.

\subsection{Some definitions and theorems}
\label{app:details}

In this section, we list some definitions and theorems that we glossed over in the main text. They can also be found in mathematics textbooks such as Ref.~\onlinecite{farenick2000algebras}.

\begin{definition} \label{def:algebra}
    An \textit{algebra} is a complex vector space $A$ equipped with associative multiplication and a multiplicative identity 1, such that
    \begin{align*}
        (x+y)z&=xz+yz,\\
        z(x+y)&=zx+zy,\\
        (\lambda x)(\mu y)&=(\lambda\mu)(xy),
    \end{align*}
    for all $x,y,z\in A$, and $\lambda,\mu\in\mathbb{C}$. An \textit{involution} is an antilinear map $x\mapsto x^*$ on $A$ such that $1^*=1$, $x^{**}=x$ and $(xy)^*=y^*x^*$ for all $x,y\in A$. The involution is \textit{positive} if $x^*x\neq0$ for all $x\neq0$.
\end{definition}

For a semisimple algebra $A$ in $(2+1)$D, the involution is defined on elementary operators by replacing anyons $a,b,c$ with their respective antiparticles $a^*,b^*,c^*$, and extended to $A$ antilinearly, i.e.\ $(\lambda x)^*=\lambda^* x^*$ where $\lambda\in\mathbb{C}$, $x\in A$ and $\lambda^*$ is the complex conjugate of $\lambda$. In the examples in this paper, all anyons are self-dual, so the involution acts trivially on the elementary operators. We can check explicitly for chiral Ising that this map is indeed an involution and is positive. Note that this check is performed manually on elementary operators for the definition of involution, and on an arbitrary operator for positivity. We cannot trivialize this check by identifying the operators with block-diagonal matrices, which would require Theorem~\ref{thm:structure}. Although the check is tedious, we do not know an easier method.

In an algebra, the structures that can be quotiented out are called ideals.
\begin{definition} \label{def:ideal}
    A subset $I\subset A$ is an \textit{ideal} if $I$ is a vector subspace of $A$ and for all $r\in I$, $x\in A$, we have $rx\in I$, $xr\in I$. In the presence of an involution, an ideal $I\subset A$ is \textit{involutive} if it is closed under the involution.
\end{definition}
Basically, an involutive ideal is a set of elements that can be identified with 0 consistently, since if $r$ is identified with 0 then so are $r^*$, $rx$ and $xr$ for all $x\in A$. If $I$ is an involutive ideal, then the quotient algebra $A/I$ is defined in the same way as for quotients of vector spaces. If $A$ is finite dimensional, then $A/I$ is also an algebra with positive involution (positivity is a consequence of Theorem~\ref{thm:structure}). When we reduced $A$ to $A_0$ in Section~\ref{sec:chiral}, we found relations among the elementary operators from physical argument, generated an ideal $I$ from the relations, and then took the quotient $A/I$. Here, if $\Omega\subset A$ is a subset, e.g.\ $\Omega=\{\omega_1,\omega_2\}$, then the ideal generated by $\Omega$ is written as \[\left<\omega_1,\omega_2\right>_{\text{id},\, A}=\{x_1\omega_1y_1+x_2\omega_2y_2\,|\,x_i,y_i\in A\},\] where the subscript $A$ indicates the overall algebra. In other words, the ideal generated by $\Omega$ is the smallest ideal of $A$ containing $\Omega$, as we need to multiply $\omega_i$ on both the left and the right, and then take linear combinations to make it an ideal. In all of the physical examples in this paper, such ideals happen to be involutive. When it is clear from context, we will drop the word ``involutive'' and simply say ``ideal''.

The fact that matrix algebras do not have non-trivial ideals can be summarized as follows:

\begin{definition}\label{def:simple}
    An algebra $A_0$ is \textit{simple} if its only (not necessarily involutive) ideals are $\{0\}$ and $A_0$ itself.
\end{definition}

\begin{lemma}\label{lemma:mat_simple}
    A finite dimensional algebra is simple if and only if it is a matrix algebra.
\end{lemma}

Note that the notions of simplicity and semisimplicity (Definition~\ref{def:ssimple}) do not rely on an involution. The following theorem relates semisimple algebras to algebras with positive involution:
\medskip
\begin{thm}\label{thm:structure}
    Let $A$ be a finite dimensional algebra with positive involution. Then $A$ is semisimple, and can be written in the form of \eqref{eq:ssimple_def} where the involution acts as Hermitian conjugation of matrices.
\end{thm}
This is why positivity of the involution is important, and the theorem fails if the involution is not positive (see example in Appendix~\ref{app:examples}). The ideals of a semisimple algebra \eqref{eq:ssimple_def} are of the form $A_{i_1}\oplus\cdots\oplus A_{i_k}$ where $1\le i_1<\cdots < i_k\le m$. In other words, to write down an ideal $I$ of $A$, we simply throw away some of the summands in \eqref{eq:ssimple_def} and keep the rest. Therefore, to make the quotient $A/I$ simple, we need to throw away precisely one $A_i$ and put the rest into the ideal $I$, and $A/I$ is isomorphic to this $A_i$.

To generate an ideal from relations, we need to use the primitive central projectors $\{P_i\}$. Suppose we want an ideal $I=\left<\{x_k\}\right>_{\text{id},\,A}$ where $\{x_k\}$ are some general elements. Let \[S=\{i\,|\,P_ix_k\neq0 \text{ for some } k\}.\] Then we have \[I=\bigoplus_{i\in S}A_i,\ A/I=\bigoplus_{i\notin S}A_i=\left(\sum_{i\notin S}P_i\right)A.\] The proof is straightforward, and the idea is that if $x_k$ has a non-trivial component in some $A_i$, then the entirety of $A_i$ must be in $I$. We can view this statement as a more general version of \eqref{eq:trick}.

Finally, we have a more rigorous version of Lemma~\ref{thm:center}:

\begin{lemma}\label{thm:center2}
    Let $B$ be a finite dimensional simple algebra with positive involution, $N$ an abelian, involutive subalgebra of $B$, and $N'$ the commutant of $N$. Then we have $Z(N')=N$.
\end{lemma}
This can be derived from the so-called von Neumann Bicommutant Theorem:
\begin{thm}
    Let $B$, $N$ and $N'$ be as in Lemma~\ref{thm:center}, and $N''$ the commutant of $N'$. Then we have $N''=N$.
\end{thm}
Using this theorem, we have $N''=N\subset Z(N')\subset N''$, so $N=Z(N')$.

When discussing Definition~\ref{def:Cartan}, we mentioned that a Cartan subalgebra must satisfy an extra condition. Here is a rigorous definition of a Cartan subalgebra:
\begin{definition}\label{def:Cartan2}
    A subalgebra $C$ of an algebra $A$ is a \textit{Cartan subalgebra} if it is abelian, diagonalizable and maximal. Diagonalizable means that every $x\in C$ is diagonalizable in its (faithful) block-diagonal matrix representation; maximal means that if any subalgebra $C'\subset A$ is abelian and diagonalizable and $C\subset C'$, then $C'=C$.
\end{definition}
Diagonalizability can also be characterized intrinsically: An element $x\in A$ is diagonalizable if and only if its minimal polynomial has distinct linear factors\cite{horn13}. This statement can be used to show that the Cartan subalgebras we chose for 1-F Ising and Ising cage-net are indeed diagonalizable, since their operators all satisfy the polynomial $t^2-1=(t+1)(t-1)$, which has distinct linear factors $(t+1)$ and $(t-1)$. Diagonalizability is needed for Lemma~\ref{thm:Cartan} to hold since e.g.\ the subalgebra of $\text{Mat}_4$ consisting of elements of the form
\[
\begin{pmatrix}
a & 0 & b & c\\
& a & d & e\\
& & a & 0\\
& & & a
\end{pmatrix}
\]
is abelian, contains non-diagonalizable elements, and has dimension 5.

\subsection{Examples of non-semisimple algebras}
\label{app:examples}

In this section, we give three examples of non-semisimple algebras and thus highlight the premises of Theorem~\ref{thm:structure}.

(1) Let $A\subset\text{Mat}_2$ be the algebra of $2\times2$ upper triangular matrices. Since $\dim(A)=3$, if $A$ is semisimple then it must be $3\text{Mat}_1$. However, this implies that $A$ is abelian, which is false. Therefore, $A$ is not semisimple. Intuitively, this can be understood as due to the lack of an involution, since $A$ is not closed under Hermitian conjugation.

(2) Let $A$ be the involutive algebra generated by two formal elements 1 and $a$, where 1 is the multiplicative identity, $a^2=0$ and $a^*=a$. This involution is not positive, so Theorem~\ref{thm:structure} does not apply here. Indeed, since $\dim(A)=2$, if $A$ is semisimple then it must be $2\text{Mat}_1$. However, we have an element $a\neq0$, $a^2=0$, but there is no such element in $2\text{Mat}_1$. Therefore, $A$ is not semisimple.

(3) Let $V$ be a complex vector space, possibly infinite dimensional. The \textit{tensor algebra} over $V$ is \[T(V)=\bigoplus_{k=0}^\infty V^{\otimes k},\] where $V^{\otimes 0}=\mathbb{C}$. The multiplication is formal, i.e.\ if $x\in V^{\otimes m}$ and $y\in V^{\otimes n}$ then $xy\in V^{\otimes(m+n)}$. The tensor algebra is always infinite dimensional regardless of $\dim(V)$, so Theorem~\ref{thm:structure} also does not apply here. Indeed, $A$ is semisimple if and only if $V=\{0\}$ (we allow infinite direct sum in Definition~\ref{def:ssimple}). Suppose, for example, that $V$ is spanned by a single element $a$. Then the quotient \[T(V)/\left<a^2\right>_{\text{id},\,T(V)}\] is precisely the $A$ in the previous example, which is not semisimple. However, we know that a quotient of a semisimple algebra is also semisimple. Therefore, $T(V)$ is not semisimple. Since semisimplicity does not rely on an involution, here we do not need to assign an involution to $T(V)$ even though we could.

Incidentally, the finite dimensional semisimple algebra $A$ in the operator algebra approach discussed in this paper can also be viewed as a quotient $T(V)/K$ of the tensor algebra. Here, $V$ is the formal vector space over the elementary operators, and $K$ is the ideal generated by multiplication rules which themselves are due to fusion rules, $F$-symbols and $R$-symbols.

\subsection{Matrix representation of simple algebra}
\label{app:matrix}

In this section, we answer the following question: Given an abstract finite dimensional simple algebra $A_0$ with positive involution, how do we find a matrix representation for it? Of course we have an isomorphism $\rho_n:A_0\to\text{Mat}_n$ for some $n$, such that the involution on $A_0$ maps to Hermitian conjugation on $\text{Mat}_n$. However, we want to determine $\rho_n$ while only assuming knowledge of the structure constants $f^\gamma_{\alpha\beta}$ with respect to some basis $\{v_\alpha\}$, as defined in \eqref{eq:structure_const}, as well as the action of the involution. This will lead to the representation \eqref{eq:matrix3} of chiral Ising operators without prior knowledge.

In our construction of $\rho_n$, we will make several claims without proof, and the proofs can be found in Ref.~\onlinecite{farenick2000algebras}. To start with, we solve the following set of linear and quadratic equations in the variables $\varepsilon_\alpha,\lambda_\alpha\in\mathbb{C}$:
\begin{equation}\label{eq:rep_start}
    \begin{aligned}
        \varepsilon^*&=\varepsilon,\\
        \varepsilon^2&=1,\\
        \varepsilon v_\alpha \varepsilon&=\lambda_\alpha \varepsilon \text{ for all }\alpha,
    \end{aligned}
\end{equation}
where $\varepsilon=\sum_\alpha \varepsilon_\alpha v_\alpha$. We claim that \eqref{eq:rep_start} always has solutions. In fact, if $n>1$ then there are many solutions, in which case we choose one solution. We can think of $\varepsilon$ as the elementary matrices whose only non-zero entry is the $(1,1)$ entry, which is 1. The variables $\lambda_\alpha$ will be of no use for us.

Let $V$ be the vector space spanned by $\{v_\alpha \varepsilon\}$. We claim that $\dim(V)=n$ even though we defined it as the span of $n^2$ elements. Clearly $V$ is closed under left multiplication by $A_0$, and indeed it is the vector space that affords the representation $\rho_n$ of $A_0$. Practically, we may reduce the overcomplete set $\{v_\alpha \varepsilon\}$ to obtain a basis for $V$. We want an inner product $\left<x,y\right>$ for all $x,y\in V$, which then defines Hermitian conjugation of matrices. By the definition of $V$, there exist $a,b\in A_0$ (not unique) such that $x=a\varepsilon$, $y=b\varepsilon$. By \eqref{eq:rep_start}, we have \[x^*y=\varepsilon a^* b\varepsilon=\lambda\varepsilon\] for some $\lambda\in\mathbb{C}$. Since $\varepsilon\neq0$, this $\lambda$ does not depend on the choice of $a,b$. We define $\left<x,y\right>=\lambda$, and we claim that this is an inner product.

The Hermitian conjugation derived from this inner product is compatible with the involution on $A_0$. 
This is because for all $z\in A_0$ and $x,y\in V$, we have \[\left<x,z^*y\right>\varepsilon=x^* z^* y\varepsilon=(zx)^* y\varepsilon=\left<zx,y\right>\varepsilon=\left<x,z^\dagger y\right>\varepsilon,\] which implies $z^*=z^\dagger$. Therefore, the action of $A_0$ on $V$ by left multiplication serves as a representation $\rho_n$.

\section{GSD of the minimal Ising cage-net}
\label{app:small_size}

In this appendix, we use the Hamiltonian \eqref{eq:H} to calculate the GSD of Ising cage-net with system size $L_x=L_y=L_z=1$ in terms of doubled Ising layers. This calculation does not involve the operator algebra approach and therefore serves as an independent check of \eqref{eq:IsingCN_GSD_formula} for the minimal system size. Indeed, we find $\text{GSD} = 144$ in agreement with \eqref{eq:IsingCN_GSD_formula}.

We start with doubled Ising on a minimal trivalent lattice, and then apply the results to Ising cage-net.

\subsection{Doubled Ising on minimal lattice}

\begin{figure}
    \centering
    \begin{tikzpicture}[baseline={([yshift=-.5ex]current bounding box.center)}, every node/.style={scale=1}]

        \pgfmathsetmacro{\lineW}{0.6}
        \pgfmathsetmacro{\ClineW}{1}
        \pgfmathsetmacro{\TlineW}{2}
        \pgfmathsetmacro{\radius}{1}
        \pgfmathsetmacro{\Squareradius}{0.5*\radius}
        
        \draw[] (\Squareradius,0) node[] (SquareEast) {};
        \draw[] (-\Squareradius,0) node[] (SquareWest) {};
        \draw[] (0,\Squareradius) node[] (SquareNorth) {};
        \draw[] (0,-\Squareradius) node[] (SquareSouth) {};
            
        \begin{scope}
            \draw [line width=\lineW] (SquareEast.center) -- ++(\radius, 0);
            \draw [line width=\lineW] (SquareWest.center) -- ++(-\radius, 0);
            \draw [line width=\lineW] (SquareNorth.center) -- ++(0,\radius);
            \draw [line width=\lineW] (SquareSouth.center) -- ++(0,-\radius);
            \draw [line width=\lineW] (SquareNorth.center) -- (SquareWest.center);
            \draw [line width=\lineW] (SquareNorth.center) -- (SquareEast.center);
            \draw [line width=\lineW] (SquareSouth.center) -- (SquareWest.center);
            \draw[line width = \lineW] (-1.5*\radius,-1.5*\radius) rectangle (1.5*\radius,1.5*\radius);
        \end{scope}
        
        \node[fill = white,minimum width = 1pt, inner sep = 2 pt] at (\radius,0) {$a$}; 
        \node[fill = white,minimum width = 1pt, inner sep = 2 pt] at (-\radius,0) {$a$}; 
        \node[fill = white,minimum width = 1pt, inner sep = 2 pt] at (0,\radius) {$b$}; 
        \node[fill = white,minimum width = 1pt, inner sep = 2 pt] at (0,-\radius) {$b$}; 
        \node[fill = white,minimum width = 1pt, inner sep = 2 pt] at (-0.5*\Squareradius,0.5*\Squareradius) {$c$}; 
        
        \begin{scope}
            \draw [line width=\ClineW, draw = blue, rounded corners = 5pt] (-0.2*\radius,\radius+\Squareradius) -- (-0.2*\radius,\Squareradius) -- (-\Squareradius,0.2*\radius) -- (-\radius-\Squareradius,0.2*\radius);
            \draw [line width=\ClineW, draw = blue,  rounded corners = 5pt] (0.2*\radius,\radius+\Squareradius) -- (0.2*\radius,\Squareradius) -- (\Squareradius,0.2*\radius) -- (\radius+\Squareradius,0.2*\radius);
            \draw [line width=\ClineW, draw = blue,  rounded corners = 5pt] (-0.2*\radius,-\radius-\Squareradius) -- (-0.2*\radius,-\Squareradius) -- (-\Squareradius,-0.2*\radius) -- (-\radius-\Squareradius,-0.2*\radius);
            \draw [line width=\ClineW, draw = blue,  rounded corners = 5pt] (\radius+\Squareradius,-0.2*\radius)--(\Squareradius,-0.2*\radius) -- (0,\Squareradius-0.2*\radius) -- (-\Squareradius+0.2*\radius,0) -- (0.2*\radius,-\Squareradius) -- (0.2*\radius,-\Squareradius-\radius);
        \end{scope}
        
        \node[fill = white,minimum width = 1pt, inner sep = 2 pt] at (\Squareradius+0.1,\Squareradius+0.1) {$\textcolor{blue}{\bm{1}}$};
        
        \draw[-latex] (-2.5*\radius,-\radius) -- ++(0.5*\radius,0);
        \draw[-latex]  (-2.5*\radius,-\radius) -- ++ (0,0.5*\radius);
        \node [] at (-2.5*\radius-0.2,-0.5*\radius) {$y$};
        \node [] at (-2*\radius,-\radius-0.2) {$x$};
        \node [] at (2.5*\radius+0.2,0) {\text{ }};
    \end{tikzpicture}
    \captionsetup{justification=Justified}
    \caption{A minimal trivalent lattice, a state vector $\ket{abc}$, and the plaquette term $B^1_p$ (the blue 1-loop).}
    \label{fig:Logical_operators_MinimalLattice}
\end{figure}
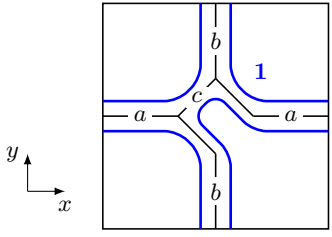

The string-net model of doubled Ising can be written on a minimal trivalent lattice (Fig.~\ref{fig:Logical_operators_MinimalLattice}). State vectors are written as $\ket{abc}$, where $a,b,c=0,1$ or 2. The subspace of the Hilbert space that satisfies the vertex terms $A_v$ has dimension 10. It is spanned by
\begin{equation}\label{eq:10_vecs}
\begin{aligned}
    w_1 &= \ket{101},\ w_2 = \ket{011},\ w_3 = \ket{110},\\
    w_4 &= \ket{121},\ w_5 = \ket{211},\ w_6 = \ket{112},\\
    w_7 &= \frac{1}{2} \ket{000} + \frac{1}{2} \ket{202} + \frac{1}{2} \ket{022} - \frac{1}{2} \ket{220} , \\
    w_8 &= \frac{1}{2} \ket{000} - \frac{1}{2}\ket{202} + \frac{1}{2} \ket{022} + \frac{1}{2} \ket{220} , \\
    w_9 &= \frac{1}{2}\ket{000} + \frac{1}{2} \ket{202} - \frac{1}{2} \ket{022} + \frac{1}{2} \ket{220} , \\
    w_{10} &= -\frac{1}{2}\ket{000} + \frac{1}{2} \ket{202} + \frac{1}{2} \ket{022} + \frac{1}{2} \ket{220}.
\end{aligned}
\end{equation}
The only nontrivial plaquette term is $B_p^1$ (Fig.~\ref{fig:Logical_operators_MinimalLattice}), which is a 1-loop that traverses each edge twice. It can also be viewed as a $\sigma$-loop (or equivalently, a $\bar\sigma$-loop) placed ``around the corners''. Using the method of \eqref{eq:corner}, we find $B_p^1=\sqrt{2}\,r$, whose eigenvalues are $\pm\sqrt{2}$. We then find
\begin{align*}
    B_p^1 w_i&=+\sqrt{2}\,w_i \text{ for } i=1,\ldots,9,\\
    B_p^1 w_{10}&=-\sqrt{2}\,w_{10}.
\end{align*}
The details of this calculation are not important and we omit it here, as $B_p^1$ does not appear in the minimal Ising cage-net since it does not commute with the condensation operators $V_{l_\mu}$. We conclude that the ground space of the minimal doubled Ising is spanned by $w_1,\ldots,w_9$.

\subsection{The minimal Ising cage-net}

The minimal Ising cage-net is obtained by condensing $\psi\bar\psi$ p-loops in three copies of minimal doubled Ising which are pairwise orthogonal. We label the states in e.g.\ the doubled Ising perpendicular to the $z$ direction by $\ket{a^z_x b^z_y c^z}$, where $a^z_x$ is on the edge in the $x$ direction, etc. The Hamiltonian consists of condensation operators $V_{l_\mu}$, vertex terms $A_v$ and a single cube term $B_c$, but with an important caveat: $B_c$ acts on a ``degenerate'' cube, whose opposite faces are identified. For example, its upper and lower faces are both proportional to \[r^z=\frac{1}{2}(1+\psi^z_x+\psi^z_y-\psi^z_x\psi^z_y).\] Since $(r^z)^2=1$, the product of these two faces is a constant. Thus $B_c$ is a constant and we can ignore it.

The subspace of the Hilbert space that satisfies the vertex terms is spanned by $w^x_i\otimes w^y_j\otimes w^z_k$, where $w^\alpha_i$ are given by \eqref{eq:10_vecs} and $i,j,k=1,\ldots,10$. According to \eqref{eq:condense_op}, in order for a state to satisfy the condensation operators $V_{l_\mu}$, we must have
\begin{equation}\label{eq:condense_min}
    (a^z_x,b^y_x),(a^x_y,b^z_y),(a^y_z,b^x_z)=(1,1)\text{ or contain no 1}.
\end{equation}
Therefore, we need to count the number of states $w^x_i\otimes w^y_j\otimes w^z_k$ that satisfy \eqref{eq:condense_min}. Up to permutation of $x$, $y$ and $z$, we have four cases:

(1) If none of the $a$'s or $b$'s (and hence $c$'s) is 1, then the states are $w^x_i\otimes w^y_j\otimes w^z_k$ where $i,j,k=7,8,9$ or 10. There are $4\times4\times4=64$ possibilities.

(2) If $(a^y_z,b^x_z)=(1,1)$ and $(a^z_x,b^y_x),(a^x_y,b^z_y)$ contain no 1, then we can take $i=2$ or 5, $j=1$ or 4, and $k=7,8,9$ or 10. There are $2\times2\times4=16$ possibilities.

(3) If $(a^z_x,b^y_x),(a^x_y,b^z_y)=(1,1)$ and $(a^y_z,b^x_z)$ contains no 1, then we can take $i=1$ or 4, $j=2$ or 5, and $k=3$ or 6. There are $2\times2\times2=8$ possibilities.

(4) If all $a$'s and $b$'s are 1, then we can take $i,j,k=3$ or 6. There are $2\times2\times2=8$ possibilities.

Summarizing these cases, we have \[\text{GSD}=64+3\times16+3\times8+8=144,\] where the factors of 3 account for permutations of $x$, $y$ and $z$. The result agrees with \eqref{eq:IsingCN_GSD_formula}.

\clearpage

\end{document}